\documentclass[a4paper,fleqn,usenatbib,useAMS]{mnras}

\usepackage{graphicx,subcaption}	% Including figure files
\usepackage{amsmath}	% Advanced maths commands
\usepackage{amssymb}	% Extra maths symbols
\usepackage{multicol}        % Multi-column entries in tables
\usepackage{bm}		% Bold maths symbols, including upright Greek
\usepackage{pdflscape}	% Landscape pages
\usepackage[encapsulated]{CJK}
\usepackage{ucs}
\usepackage[utf8x]{inputenc}
\usepackage[dvipsnames]{xcolor}
\usepackage{tikz,hyperref}

\definecolor{lime}{HTML}{A6CE39}
\DeclareRobustCommand{\orcidicon}{
	\begin{tikzpicture}
	\draw[lime, fill=lime] (0,0) 
	circle [radius=0.14] 
	node[white] {{\fontfamily{qag}\selectfont \tiny ID}};
	\draw[white, fill=white] (-0.0625,0.095) 
	circle [radius=0.007];
	\end{tikzpicture}
	\hspace{-2mm}
}

\foreach \x in {A, ..., Z}{\expandafter\xdef\csname orcid\x\endcsname{\noexpand\href{https://orcid.org/\csname orcidauthor\x\endcsname}
			{\noexpand\orcidicon}}
}
\usepackage[T1]{fontenc}
\usepackage{ae,aecompl}

\usepackage{newtxtext,newtxmath}
\title[PNe with WR CSPN – II]{Planetary nebulae with Wolf-Rayet-type central stars - II. Dissecting the compact planetary nebula M\,2-31 with GTC MEGARA}

\author[Rechy-Garc\'{i}a et al.]{J.S. Rechy-Garc\'ia\thanks{E-mail: j.rechy@irya.unam.mx}$^{1\orcidA}$, J.A.\,Toal\'{a}$^{1\orcidB}$, S.\,Cazzoli$^{2\orcidC}$, M.A.\,Guerrero$^{2\orcidD}$, L.\,Sabin$^{3\orcidE}$, V. M. A. G\'omez-Gonz\'alez$^{1\orcidF}$, \newauthor and G.\,Ramos-Larios$^{4\orcidG}$\\
$^{1}$Instituto de Radioastronom\'ia y Astrof\'isica, UNAM Campus Morelia, Apartado Postal 3-72, 58090 Morelia, Michoac\'an, Mexico\\
$^{2}$Instituto de Astrof\'\i sica de Andaluc\'\i a, IAA-CSIC, Glorieta de la Astronom\'\i a s/n, E-18008, Granada, Spain \\
$^{3}$Instituto de Astronom\'\i a, Universidad Nacional Autonoma de M\'{e}xico, Apartado Postal 877, 22800 Ensenada, B. C., Mexico\\
$^{4}$Instituto de Astronom\'\i a y Meteorolog\'\i a, CUCEI, Univ.\ de Guadalajara, Av.\ Vallarta 2602, Arcos Vallarta, 44130 Guadalajara, Mexico\\
}

\pubyear{2021}

\begin{document}
\label{firstpage}
\pagerange{\pageref{firstpage}--\pageref{lastpage}}
\maketitle

\begin{abstract}
We present a comprehensive analysis of the compact planetary nebula M\,2-31 investigating its spectral properties, spatio-kinematical structure and chemical composition using GTC MEGARA integral field spectroscopic observations and NOT ALFOSC medium-resolution spectra and narrow-band images. The GTC MEGARA high-dispersion observations have remarkable tomographic capabilities, producing an unprecedented view of the morphology and kinematics of M\,2-31 that discloses a fast spectroscopic bipolar outflow along position angles 50$^\circ$ and 230$^\circ$, an extended shell and a toroidal structure or waist surrounding the central star perpendicularly aligned with the fast outflows. These observations also show that the C\,{\sc ii} emission is confined in the central region and enclosed by the [N\,{\sc ii}] emission. This is the first time that the spatial segregation revealed by a 2D map of the C~{\sc ii} line implies the presence of multiple plasma components. The deep NOT ALFOSC observations allowed us to detect broad WR features from the central star of M\,2-31, including previously undetected broad O~{\sc vi} lines that suggest a reclassification as a [WO4]-type star. 
\end{abstract}


%%%%%%%%%%%%%%%%%%%%%%%%%%%%%%%%%%%%%%%%%%%%%%%%%%

%%%%%%%%%%%%%%%%% BODY OF PAPER %%%%%%%%%%%%%%%%%%

\section{Introduction}

Planetary nebulae (PNe) are the descendants of low- and intermediate-mass stars ($M_\mathrm{ZAMS} \lesssim 1 - 8 $~M$_\odot$), when they have ejected their hydrogen-rich envelopes through a dense and slow (10 km~s$^{-1}$) wind in the final stages of the asymptotic giant branch (AGB) phase. As the remaining stellar nucleus evolves into the post-AGB phase, it develops a fast wind \citep[$v_{\infty}\gtrsim$1000~km~s$^{-1}$;][]{Guerrero2013} that pushes and compresses the AGB material. 
It also develops a hard UV flux that photoionises the material, creating the PN \citep{Kwok2000}. %$\pm$90 km~s$^{-1}$

The rich morphological variety of PNe indicates complex material ejection processes, including the interaction between stellar winds with different degrees of symmetry and the action of high-velocity collimated outflows \citep{Balick1987,Sahai1998}. The latter would deposit momentum and kinetic energy onto the spherical nebular envelope, causing notable effects in the shaping of PNe that will finally have notably asymmetric morphologies. It is currently accepted that binary systems give birth to such bipolar structures \citep[see, for example,][and references therein]{Livio1988,Soker2001,Nordhaus2006,DeMarco2009,Zou2020,Balick2020,GS2021}.

The morphological classification of a PN or the identification of small-scale morphological components can be compromised by the image quality or depth. 
This is particularly the case for compact high-surface brightness PNe, which are presumably young PNe.  
In this case, high-dispersion spectra in the H$\alpha$ and [N~{\sc ii}] emission lines have been used to find evidence of fast collimated outflow still embedded within the nebular envelope. 
These PNe, referred as spectroscopic bipolar nebulae \citep{Rechy-Garcia2017}, allow us to examine the critical moment when collimated outflows interact with the surrounding nebular envelope. 
Morpho-kinematic studies of these nebulae are thus very valuable to assess the effects of collimated outflows in the formation of asymmetric PNe.

\begin{figure*}
	\includegraphics[width=0.95\linewidth, trim=0 120 0 0cm]{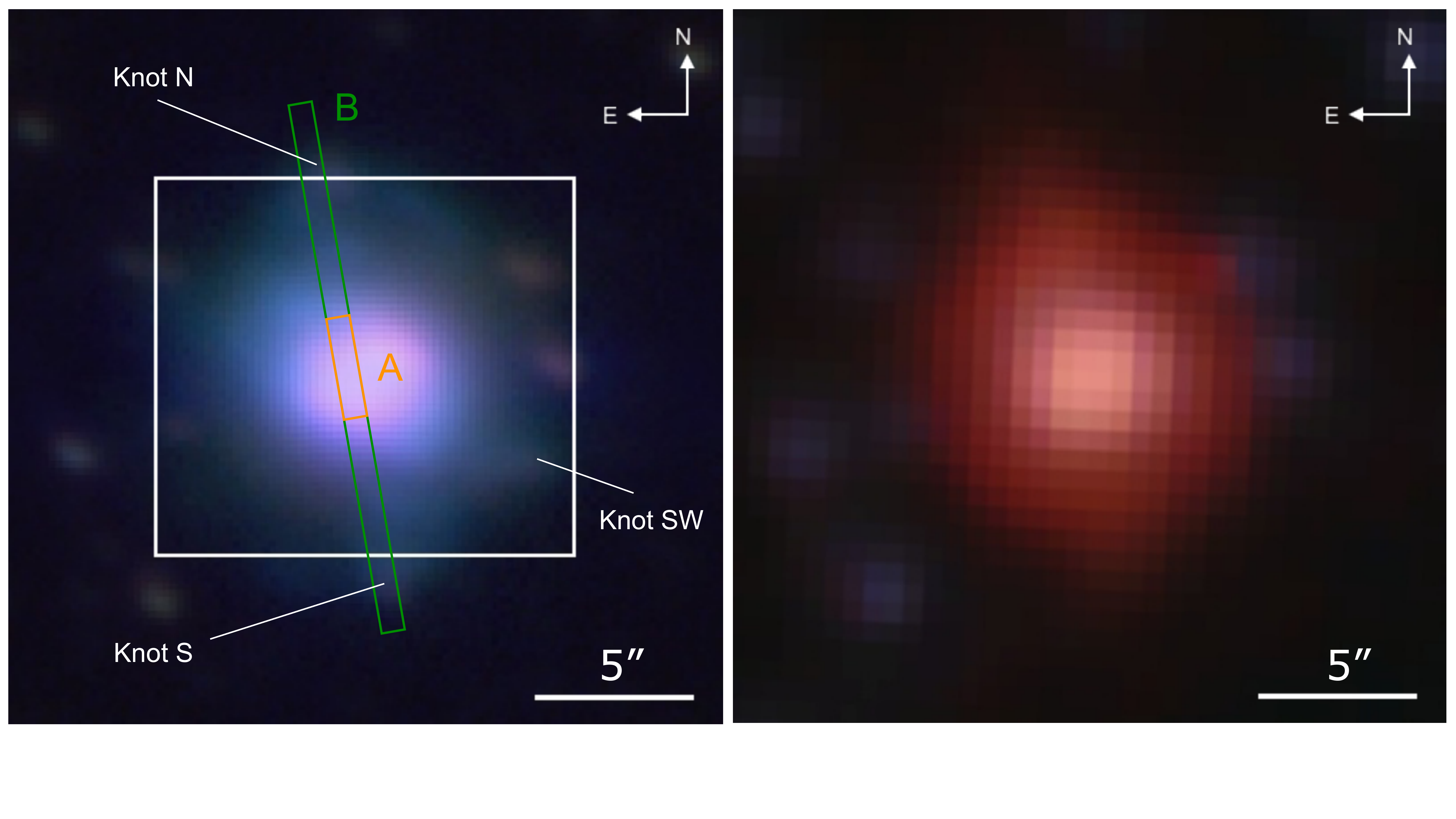}
\caption{{\it (left)} NOT ALFOSC colour-composite image of M\,2-31 in H$\alpha$ and $r'$-SDSS (green), [N~{\sc ii}] (red) and [O~{\sc iii}] (blue). The extraction regions used for the NOT ALFOSC slits are illustrated with an orange region for the CSPN (labeled as A) and two green rectangular region for the nebula (labeled as B). The area covered by the GTC MEGARA IFU is marked by a white rectangle. Some features are labeled (see text for details). {\it (right)} {\it Spitzer} colour-composite image in the 4.5 $\mu$m (blue), 5.8 $\mu$m  (green) and 8 $\mu$m (red) IRAC bands.
}
\label{fig:Componentes}
\end{figure*}

The PN M\,2-31 (PN\ G006.0-03.6) belongs to this group of spectroscopic bipolar nebulae.  
M\,2-31 is a compact PN with an angular size of 4\farcs0 $\times$ 3\farcs7 (see Figure~\ref{fig:Componentes}) located at a distance of 5.9$\pm$1.7 kpc \citep{Frew2016} towards the direction of the Galactic Centre \citep{Gathier1983}. 
Long-slit echelle spectroscopic observations of M\,2-31 have unveiled faint features with expansion velocities reaching $\pm$90 km~s$^{-1}$ that are indicative of a collimated bipolar outflow, and a faint and seemingly slightly tilted elliptical shell that surrounds a number of bright knots or blobs \citep{AkrasLopez2012}.

Previous works reported that M\,2-31 hosts a [Wolf-Rayet] C-rich central star (CSPN) with a [WC4-6] spectral type  \citep{Tylenda1993,AckerNeiner2003}. The nebula itself has a dual-dust chemistry \citep{DelgadoIngladaRodriguez2014}. 
Such objects, first discovered by \citet{Waters1998}, are described as PNe with strong polycyclic aromatic hydrocarbons (PAHs) and silicate features. Although associated with [WC]-type CSPNe, this phenomenon is not only restricted to this class of CSPNe \citep{PereaCalderon2009}. The electron density ($n_\mathrm{e}$) reported for this nebula is 7250 cm$^{-3}$, which is suggestive of an early evolutionary stage. Its abundance discrepancy factor (ADF) is 2.42 \citep{GarciaRojas2018}, a mean value for PNe \citep[see][]{Wesson2018}.

Here we present high spatial-resolution optical images, intermediate-dispersion long-slit spectroscopic data and high-dispersion integral field spectroscopic (IFS) observations of M\,2-31 to investigate its morphology, kinematics, 3D physical structure, physical conditions, chemical abundances and the properties of its CSPN.  
High-dispersion IFS observations of PNe provide 2D spatial information on their kinematics  allowing to search for hidden spatio-kinematic components that can be missed using long-slit echelle observations.  
The capabilities of this tomographic techniques have been demonstrated by our group for HuBi\,1, where a shell-like structure has been revealed for its born-again ejecta \citep{RechyGarcia2020}, and for NGC\,2392, where it has been possible to obtain an image of its jet for the first time, providing detailed information on its morphology and kinematics \citep{Guerrero2021}.

We next describe the observations in Section~2 and provide the results in Section~3.  
A discussion is presented in Section~4 and the main results and conclusions are summarized in Section~5.

\section{observations}

\subsection{Imaging}

Narrow-band H$\alpha$, [N~{\sc ii}], and [O~{\sc iii}] images of M\,2-31 were acquired on 2020 July 29 using ALFOSC (Alhambra Faint Object Spectrograph and Camera)\footnote{\url{http://www.not.iac.es/instruments/alfosc/}} at the 2.56 m Nordic Optical Telescope (NOT) of the Observatorio de El Roque de los Muchachos (ORM, La Palma, Spain). 
One 300 s exposure was obtained through the H$\alpha$ filter, three 300 s exposures through the [N~{\sc ii}] filter, 
five 600 s exposures through the [O~{\sc iii}] filter and five 30 s exposures in $r'$-SDSS using the E2V 231-42 2k$\times$2k CCD with plate scale of 0\farcs211 pix$^{-1}$. In all cases, a dithering of a few arcsecs was applied between individual exposures to improve the quality of the final image. The seeing, according to the FWHM of stars in the field of view, was 1.2 arcsec. 
The individual exposures were bias subtracted and flat-fielded using twilight sky frames, and then aligned and combined to remove cosmic rays using standard {\sc iraf} routines \citep{Tody1993}. A colour-composite picture of M\,2-31 using these images is presented in the left panel of Figure~\ref{fig:Componentes}.

For comparison with the optical images we also present a colour-composite mid-IR picture of M\,2-31 in the right panel of  Figure~\ref{fig:Componentes} created using {\it Spitzer} Infrared Array Camera (IRAC) observations \citep{Fazio2004}. 
The IRAC observations were obtained on 2007 October 17 for a total exposure time of 280~s as part of the Program ID.\ 40115 (PI: G\,Fazio). These were retrieved from the NASA/IPAC Infrared Science Archive\footnote{\url{https://irsa.ipac.caltech.edu/frontpage/}}. 

\subsection{Integral field spectroscopy}

IFS observations were obtained on 2020 August 19 with the Multi-Espectr\'ografo en GTC de Alta Resoluci\'on para Astronom\'ia \citep[MEGARA;][]{GilDePaz2018} at the Gran Telescopio Canarias (GTC) of the ORM.  
The Integral Field Unit (IFU) mode was used.  
It covers a field of view (FoV) of 12\farcs5$\times$11\farcs3 on the sky with 567 hexagonal spaxel of maximal diameter 0\farcs62, which makes it ideal for the study of compact PNe. 
The observations were obtained with the High-Resolution Volume Phased Holographic (VPH 665-HR) covering the 6405.6-6797.1 \AA\ wavelength range with spectral resolution of R = 18,700 (i.e., $\simeq16$ km~s$^{-1}$). 
Three exposures of 300 s were obtained with a seeing of 0\farcs7 during the observations. The white rectangle in the left panel of Figure~\ref{fig:Componentes} is the MEGARA IFU FoV.

The MEGARA raw data were reduced using the Data Reduction Cookbook provided by Universidad Complutense de Madrid \citep{Pascual2019}. 
The pipeline allows us to subtract the sky and bias contributions, apply the flat field correction, and perform wavelength calibration, spectra tracing and extraction. 
The sky subtraction is done using 56 ancillary fibres located $\approx2\farcm0$ from the IFU center.
The flux calibration was performed using observations of the spectro-photometric standard HR\,7596 obtained immediately after those of M\,2-31.

The analysis of the data was performed considering squared spaxels of size of 0\farcs215 per pixel getting a final cube with dimensions of 52$\times$58$\times$4300 using the regularization grid task {\it megararss2cube.}\footnote{
Task developed by J.\,Zaragoza-Cardiel available at \url{https://github.com/javierzaragoza/megararss2cube}.
}

\subsection{Long-slit spectroscopy}

Long-slit spectroscopic observations were obtained using also ALFOSC at the NOT of the ORM (PI: M. A. Guerrero). 
Medium-resolution spectra were obtained on 2020 July 29 using the E2V CCD 231-42 2k$\times$2k camera with the Grism \#7, which provides a resolution of $R=1000$.
This configuration covers the 3650–7110 \AA\ wavelength range with a plate scale of 0\farcs211 pix$^{-1}$. 
The slit width was set at 0\farcs75 with a position angle (PA) of 10$^{\circ}$. 
Two exposures of 450 s were obtained through this slit during the observations.

The spectra were analysed following {\sc iraf} standard routines. The wavelength calibration was performed using HeNe lamps. In order to study the chemical abundance and the central star of M\,2-31, we extracted spectra from different regions defined on the slit using the {\sc iraf} task \emph{apall}. Two green rectangular regions are shown in Figure~\ref{fig:Componentes} which represent the extractions carried out for the study of the chemical abundances of the nebula labeled with letter B, and one orange region labeled with letter A is the extraction region used for study the CSPN.
The spectrum extracted from region A with an aperture size of 3\farcs2 was used to study the CSPN, whereas the spectrum extracted from region B with an aperture size of 17\farcs1 was used to study the PN. 
We first extracted the 1D spectra of the CSPN (region A) and nebula (region B) tracing the stellar continuum along the 2D spectrum. 
The 1D spectrum of the CSPN was then subsequently subtracted from the 1D spectrum of the aperture B to obtain a pure nebular spectrum of M\,2-31 with a net aperture size of 13\farcs9.
These 1D spectra are presented and discussed in subsections~\ref{sec:spec_cspn} and \ref{sec:spec_neb}.

\section{RESULTS}

\subsection{Morphology}

The NOT colour-composite picture of M\,2-31 in the left panel of Figure~\ref{fig:Componentes} 
shows bright [N~{\sc ii}] emission in the central zone within a region of size 2\farcs0$\times$1\farcs7. 
This region is surrounded by a fainter shell with an angular radius $\approx$7\farcs3. 
Several knots can be identified at different positions: 
two towards the North and South (hereinafter labeled as Knot N and Knot S, respectively) at PA = 10$^{\circ}$ and PA = 190$^{\circ}$ with radial distance $\approx$ 6\farcs0, and another one towards the South-West (hereinafter labeled as Knot SW) located at PA = 240$^{\circ}$ with radial distance $\approx$ 5\farcs3. The N and S knots seem to give the impression that M\,2-31 has an elongated shape with its symmetry aligned with these knots, but we will demonstrate in this paper that this is not the case.

The right panel of Figure~\ref{fig:Componentes} shows a colour-composite \emph{Spitzer} image that combines 4.5 $\mu$m (blue), 5.8 $\mu$m (green) and 8 $\mu$m (red) IRAC images. 
The morphology unveiled by this mid-IR picture is consistent with the optical image, with a bright inner structure surrounded by a fainter shell. The N and S knots of the optical images are hinted in the \emph{Spitzer} IRAC 8 $\mu$m image, resulting in a somewhat bipolar shape in the mid-IR for M\,2-31. No obvious mid-IR emission is detected from the SW knot.

\subsection{Kinematics}

\begin{figure*}
\begin{center}
	\includegraphics[width=1.0\columnwidth, trim=0 0 0 0cm]{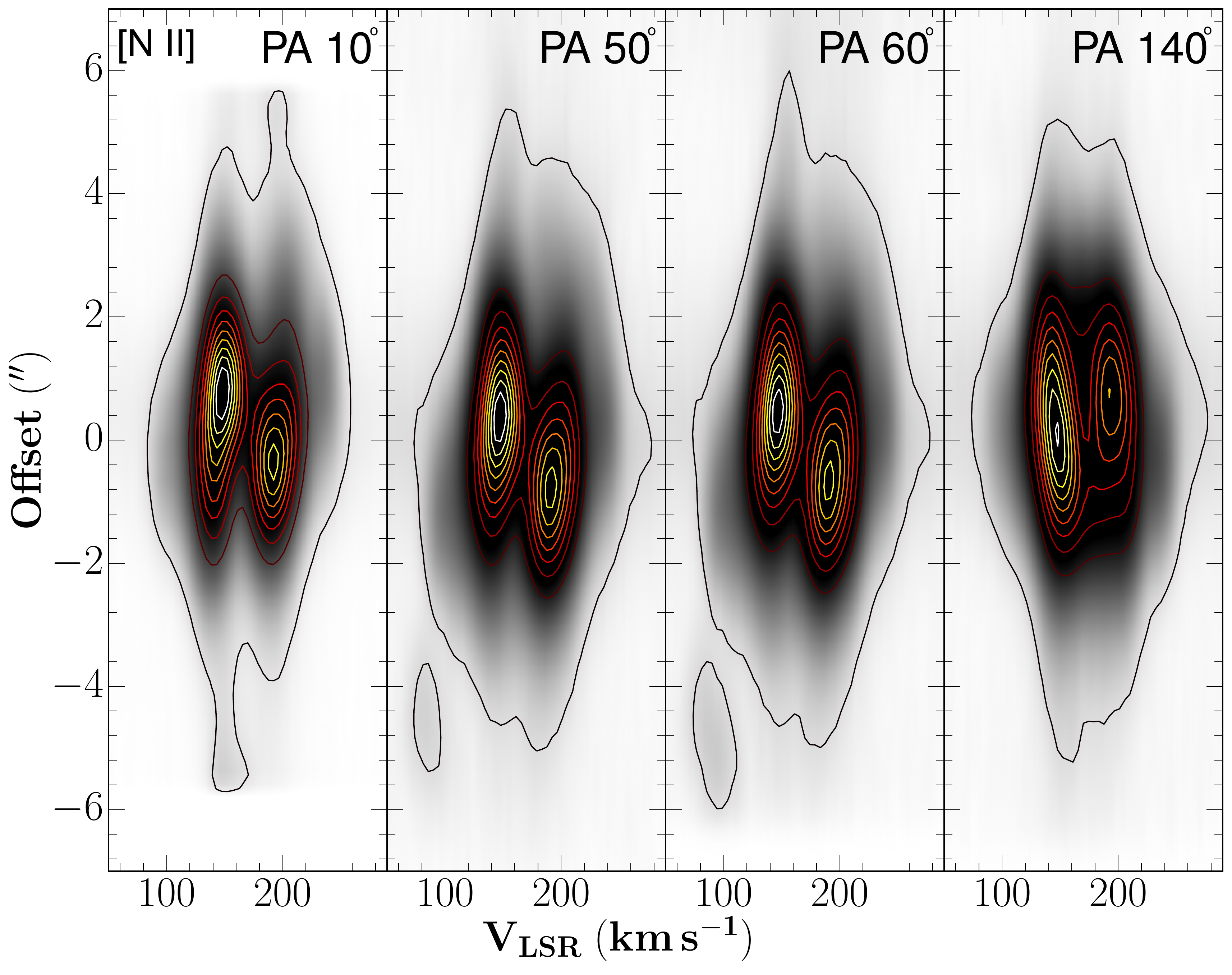}
	\includegraphics[width=1.0\columnwidth, trim=0 0 0 0cm]{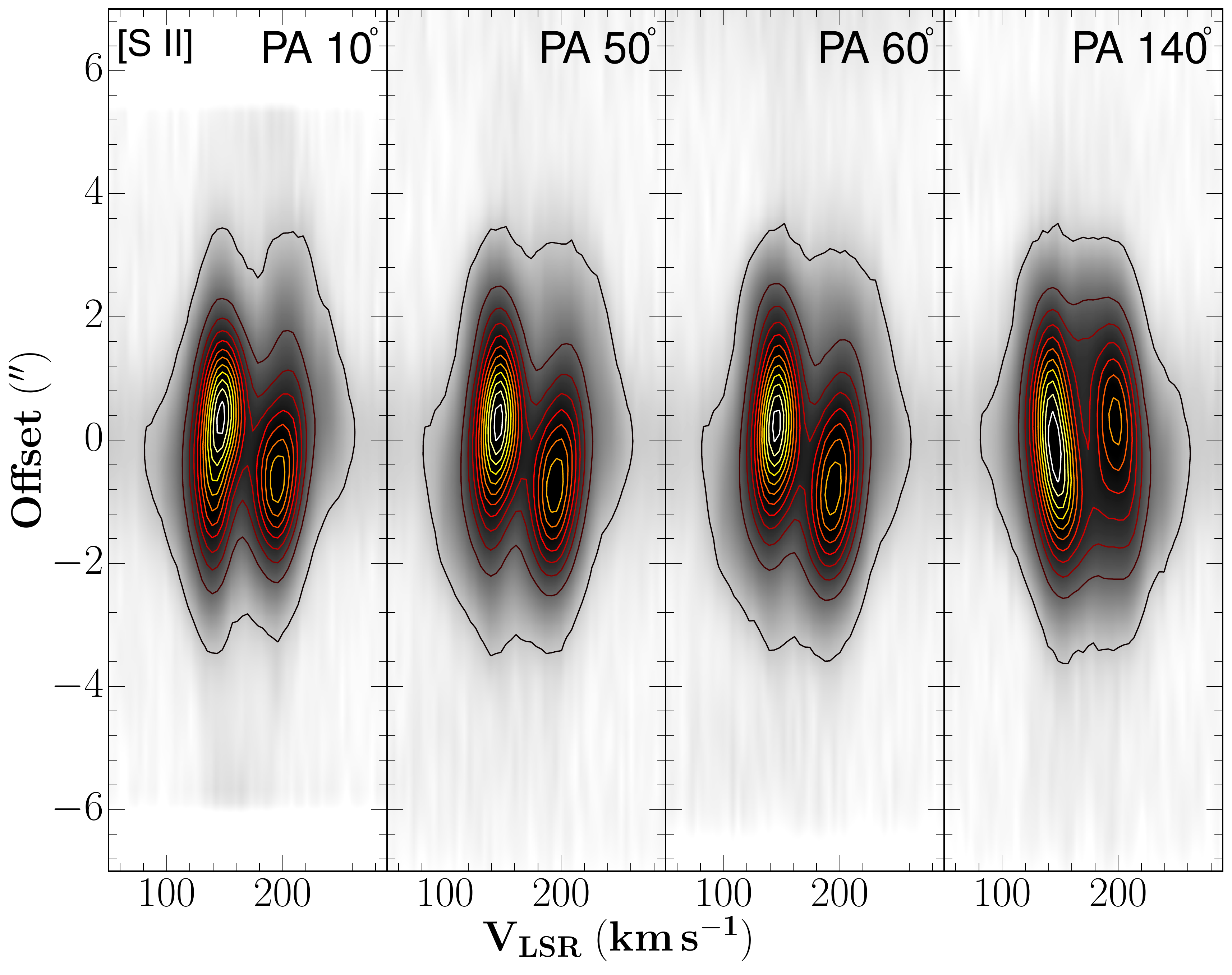}
	\includegraphics[width=1.0\columnwidth, trim=0 0 0 0cm]{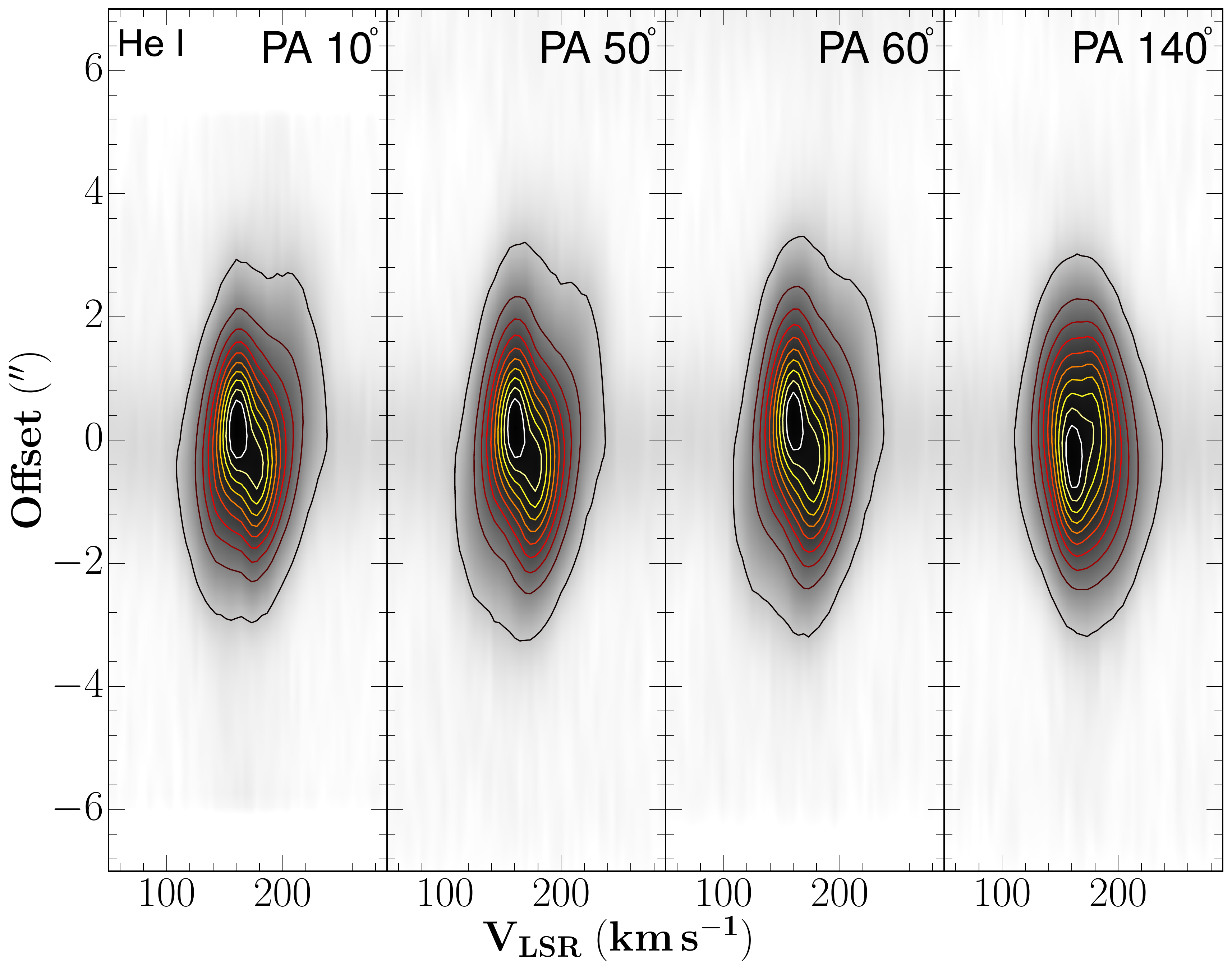}
	\includegraphics[width=1.0\columnwidth, trim=0 0 0 0cm]{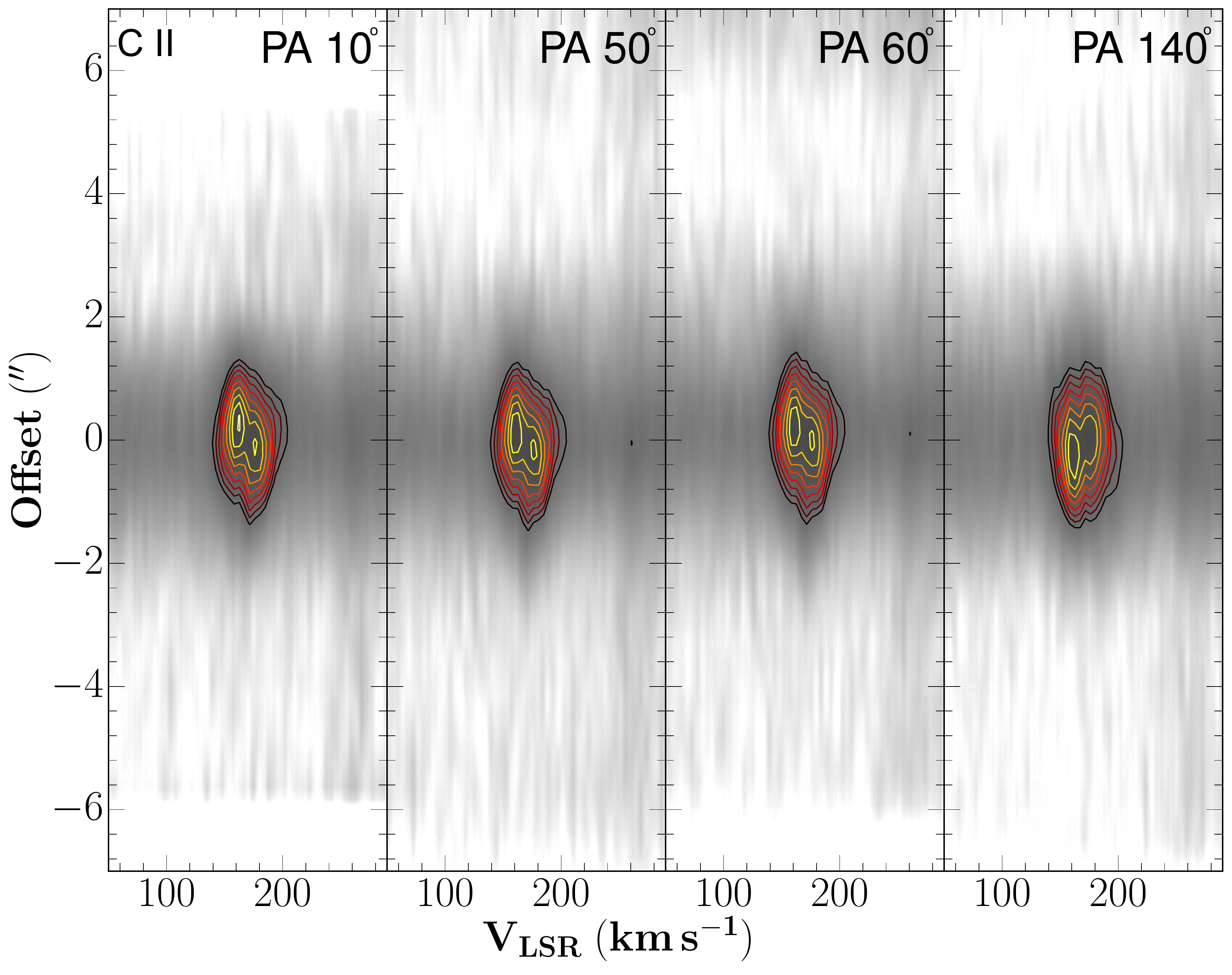}
\caption{PV maps of the [N\,{\sc ii}] $\lambda$6584 (left-top), [S\,{\sc ii}] $\lambda\lambda$6716,6731 (right-top), He\,{\sc i} $\lambda$6678 (left-bottom), and C\,{\sc ii} $\lambda$6578 (right-bottom) emission lines extracted from MEGARA pseudo-slits at PAs 10$^{\circ}$, 50$^{\circ}$, 60$^{\circ}$, and 140$^{\circ}$. Contours of different colors are used to highlight both faint and bright emissions.}
\label{fig:PVmaps}
\end{center}
\end{figure*}

The MEGARA IFS observation of M\,2-31 provides 2D information of the nebular kinematics encoded in the H$\alpha$, [N~{\sc ii}] $\lambda\lambda$6548,6584, [S~{\sc ii}] $\lambda\lambda$6716,6731, He~{\sc i} $\lambda$6678, and C~{\sc ii} $\lambda$6578 emission lines. 
A first step to analyse this information implies the identification of the kinematic signature of the morphological features spotted in Figure~\ref{fig:Componentes}.
This has been done using of the software {\sc qfitsview}\footnote{\url{https://www.mpe.mpg.de/~ott/QFitsView/}} to extract pseudo-slits in the MEGARA data cube of M\,2-31 and build position-velocity (PV) maps along the PAs of interest: 10$^{\circ}$ along the N and S knots, 50$^{\circ}$ and 60$^{\circ}$ along the SW knot, and 140$^{\circ}$ along an intermediate position.
The resultant PV diagrams are shown in Figure~\ref{fig:PVmaps}. 
According to them, the radial velocity of M\,2-31 in the LSR system is V$_\mathrm{LSR}\simeq$ 170 km~s$^{-1}$, in agreement with that reported by \citet{Schneider1983} of 168.7 km~s$^{-1}$.

Two bright blobs are visible in the [N~{\sc ii}] $\lambda$6584 and [S\,{\sc ii}] $\lambda\lambda$6716,6731 PV maps presented in Figure~\ref{fig:PVmaps} at V$_\mathrm{LSR} \simeq$ 150~km~s$^{-1}$ and $\simeq190$~km~s$^{-1}$. 
Their peaks are located within the inner $\lesssim2^{\prime\prime}$ in radius of M\,2-31, but their kinematic signature extends to distances $\sim6^{\prime\prime}$ from the center, i.e., the whole MEGARA FoV.

The PV map of the [N~{\sc ii}] emission line at PA = 10$^\circ$ discloses blobs of emission that can be attributed to the knots N and S marked in Figure~\ref{fig:Componentes}, although we note that these are just outside MEGARA's FoV.
Similarly, the [N\,{\sc ii}] PV maps extracted at PAs 50$^{\circ}$ and 60$^{\circ}$ reveal a component at a radial distance $\simeq5^{\prime\prime}$ moving at high velocity with V$_\mathrm{LSR} \simeq$ 90~km~s$^{-1}$ and with an extension of $\simeq2^{\prime\prime}$. This high-velocity component, with a velocity of 80 km~s$^{-1}$ with respect to the systemic velocity, corresponds to the SW knot labeled in Figure~\ref{fig:Componentes}. In addition, these PVs unveil the presence of two faint components very close to the central region ($\simeq0^{\prime\prime}$) whose wings extend up to  V$_\mathrm{LSR} \simeq$ 90~km~s$^{-1}$ and $\simeq$ 260~km~s$^{-1}$.

The emission revealed in the PV maps of the He\,{\sc i} and C\,{\sc ii} emission lines are fainter than that of the other emission lines (see Fig.\ref{fig:PVmaps} bottom panels). 
The He\,{\sc i} emission peaks around $<$0\farcs5 and extends up to $\lesssim3^{\prime\prime}$ from the CSPN of M\,2-31. 
The velocity difference between the He\,{\sc i} peaks is 30~km~s$^{-1}$, which is smaller than the 40~km~s$^{-1}$ velocity difference that can be inferred for the peaks of the [N~{\sc ii}] and [S~{\sc ii}] emission lines. 
Meanwhile, the C\,{\sc ii} emission peaks at $\lesssim$0\farcs3 from the center and is present only within the innermost $\sim$1\farcs2. The emission from the C~{\sc ii} line is thus the less extended among the different emission lines shown in Figure~\ref{fig:PVmaps}.

\subsection{Multi-component decomposition}
\label{sec:multi-component}

In this section we present a more detailed analysis of the kinematic structure of M\,2-31 applying to the MEGARA data an analysis method similar to that used to investigate the spatio-kinematic of the jets in NGC\,2392  \citep{Guerrero2021}. 
Briefly, this method fits the [N\,{\sc ii}] $\lambda$6584 line profile at each spaxel using multiple Gaussian components. 
The fitting is performed using the Levenberg-Marquardt least-squares fitting routine {\sc mp-fitexpr} \citep{Markwardt2009} within the Interactive Data Language ({\sc idl}) environment \citep[for more detailes see also][]{Cazzoli2020}.

\begin{figure*}
\begin{center}
\includegraphics[width=\linewidth]{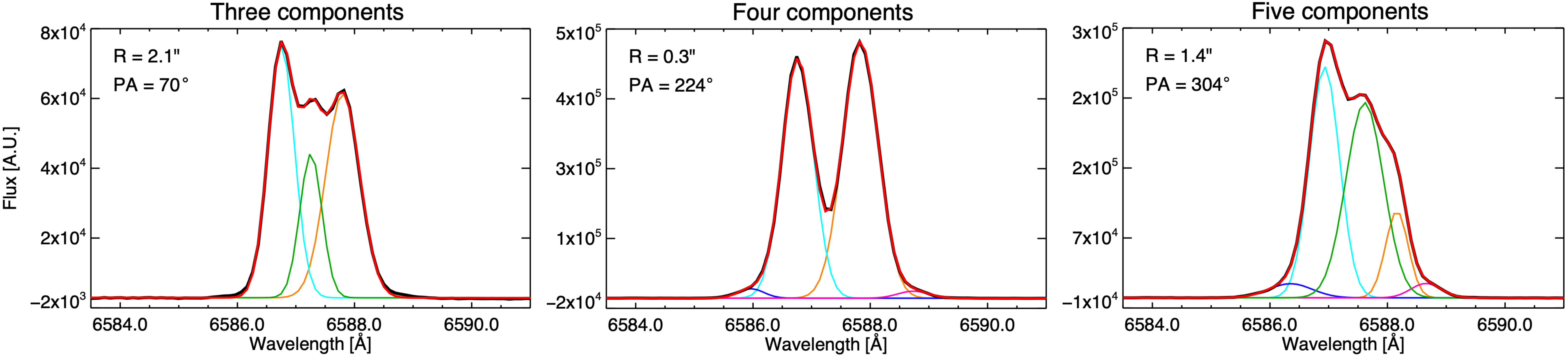}
\caption{Examples of multi-Gaussian fitting of the [N~{\sc ii}] $\lambda$6584 emission line at different spaxels of the MEGARA IFU involving three (left), four (center), and five (right) components. 
The distance to the center and position angle (PA) is labeled on the upper-right corner of each panel. 
The black line is the observed profile of the [N\,{\sc ii}] emission line. 
The blue, green, cyan, orange, and magenta lines represent the five different velocity components detected in M\,2-31 (see text), whilst the red line is the sum of velocity components. 
The residuals (not shown here) represent less than 2\% of the flux peak for the three cases.}
\label{fig:spaxel_fit}
\end{center}
\end{figure*}

\begin{figure*}
\begin{center}
\includegraphics[width=\linewidth, trim=100 20 80 0cm]{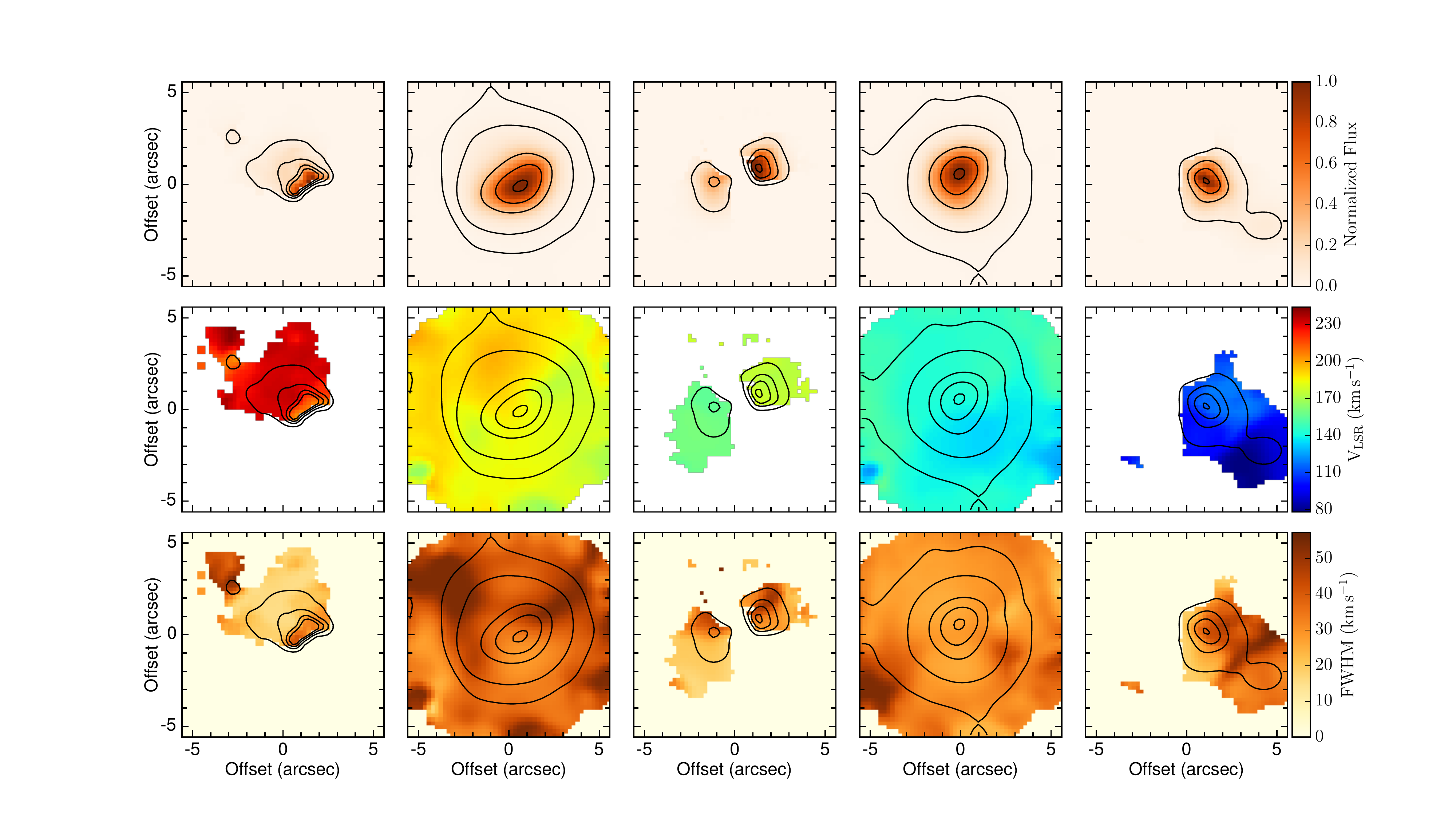}
\caption{
GTC MEGARA [N\,{\sc ii}] $\lambda$6584 normalized flux (top), velocity V$_{\rm LSR}$ (middle) and FWHM (bottom) maps of the different kinematic components of M\,2-31. 
These components are arranged in increasing velocity from left to right.  
The leftmost and rightmost columns correspond to the outflow along PA $\simeq$50$^\circ$--240$^\circ$, the second and fourth columns to the inner and outer shells of M\,2-31, and the central column to a low-velocity component almost orthogonal to the outflow described above. 
Contours derived from the normalized flux images of each component are overlaid over the velocity and FWHM maps.}
\label{fig:M2-31_MEGARA-mosaico}
\end{center}
\end{figure*}

The [N\,{\sc ii}] $\lambda$6584 emission line profiles present multiple peaks with intensities and number varying across the FoV of the MEGARA IFU. Up to five distinct kinematic components with different spatial extents and locations have been considered to fit the line profiles extracted at different spatial positions of the MEGARA data cube. Examples of those fits at different positions requiring three, four or five Gaussian components are presented in Figure~\ref{fig:spaxel_fit}.

The normalized flux (top row), velocity V$_\mathrm{LSR}$ (middle row) and FWHM (bottom row) maps of the different components in the [N\,{\sc ii}] $\lambda$6584 emission line are shown in Figure~\ref{fig:M2-31_MEGARA-mosaico}. 
These five kinematic components shown in Figure~\ref{fig:M2-31_MEGARA-mosaico} can be adscribed to three different morpho-kinematic structures.

The leftmost and rightmost columns of this figure show  components protruding from the inner core of M\,2-31 with the fastest velocities. We will refer to these two components as spectroscopic outflows.  
These are low-ionization structures detected both in the [N~{\sc ii}] and [S~{\sc ii}] emission lines. 
The red component of this spectroscopic outflow (leftmost column) is oriented with a PA $\approx$ 50$^{\circ}$, whilst the blue component (rightmost column) has a PA $\approx$ 230$^{\circ}$. 
The average V$_\mathrm{LSR}$ of the red component is 228.5~km~s$^{-1}$ and 100.9~km~s$^{-1}$ for the blue component, for a velocity difference that implies an expansion velocity of $\pm$63.8~km~s$^{-1}$ with respect to the nebula systemic velocity.

The second and fourth columns in Figure~\ref{fig:M2-31_MEGARA-mosaico} are two bright structures with average V$_\mathrm{LSR}$ of 185.3~km~s$^{-1}$ and 142.8~km~s$^{-1}$, respectively. The expansion velocity thus corresponds to 21.3~km~s$^{-1}$. 
They very likely correspond to a cavity or shell surrounding the inner regions of M\,2-31, which is visible in the optical images presented in Figure~\ref{fig:Componentes}. The second and fourth panels of Figure~\ref{fig:M2-31_MEGARA-mosaico} also reveal hints of the N and S knots at the northern and southern edges of the MEGARA FoV along PA $\simeq$10$^\circ$.

Finally, the central column of Figure~\ref{fig:M2-31_MEGARA-mosaico} discloses a pair of slowly expanding structures. 
This does not resemble a bipolar ejection, but rather a disk-like or toroidal structure that we will refer to as low-velocity component from here on. The East side of the low-velocity component has an average V$_\mathrm{LSR}$ of 158.9~km~s$^{-1}$ and the West side has an average V$_\mathrm{LSR}$ of 175.1~km~s$^{-1}$, implying an expansion velocity $\pm$8.1~km~s$^{-1}$. 
This structure, with a PA $\approx$ 110$^{\circ}$, appears to be almost orthogonal to the spectroscopic outflow.

\begin{figure}
\begin{center}
\includegraphics[width=0.61\linewidth, trim=100 20 80 0cm]{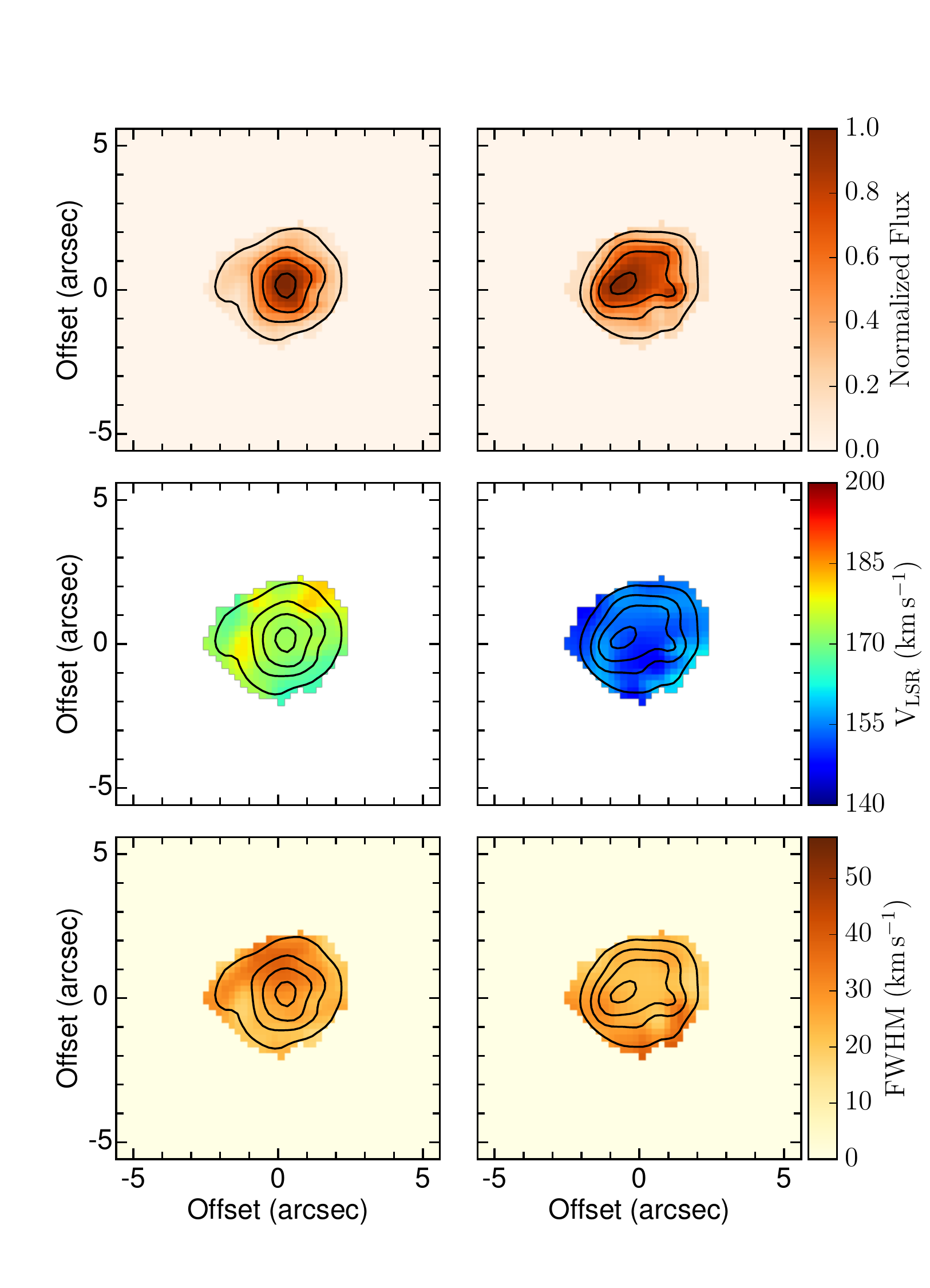}
\caption{GTC MEGARA C\,{\sc ii} $\lambda$6578 normalized flux (top), velocity V$_{\rm LSR}$ (middle) and FWHM (bottom) maps of the two kinematic components in increasing velocity from left to right.  
Contours derived from the normalized flux images of each component are overlaid over the velocity and FWHM maps.}
\label{fig:M2-31_MEGARA-mosaico_CII}
\end{center}
\end{figure}

The same line-fitting procedure was also applied to the C\,{\sc ii} $\lambda$6578~\AA\, emission line and the results are shown in Figure~\ref{fig:M2-31_MEGARA-mosaico_CII}. The overall extent of the C\,{\sc ii} emission is more compact than that exhibited by the [N\,{\sc ii}] maps. The line-fitting procedure disclosed the presence of two structures with average V$_\mathrm{LSR}$ of 186.1~km~s$^{-1}$ and 153.6~km~s$^{-1}$, respectively. This implies an expansion velocity $\simeq$16.3 km~s$^{-1}$, which is smaller than that derived for the two brightest [N~{\sc ii}] components. Furthermore, the FWHM of the C~{\sc ii} components (from 10 to 30 km~s$^{-1}$) are narrower than that of [N\,{\sc ii}] in the same regions (from 30 to 50 km~s$^{-1}$).

\begin{figure}
\begin{center}
\includegraphics[width=0.77\linewidth]{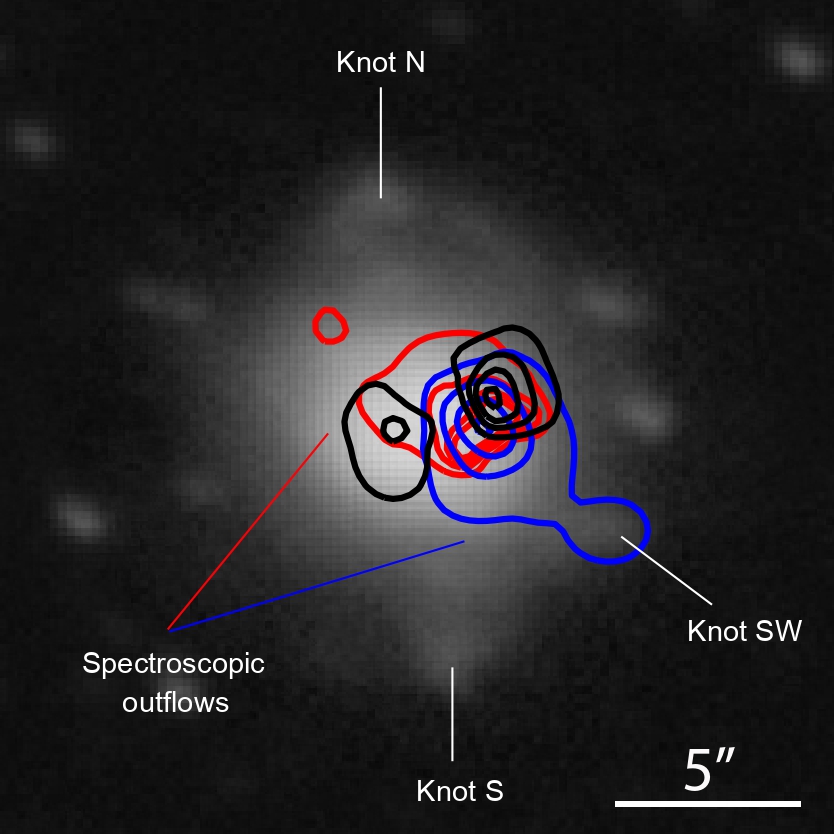}\\
\includegraphics[width=0.77\linewidth]{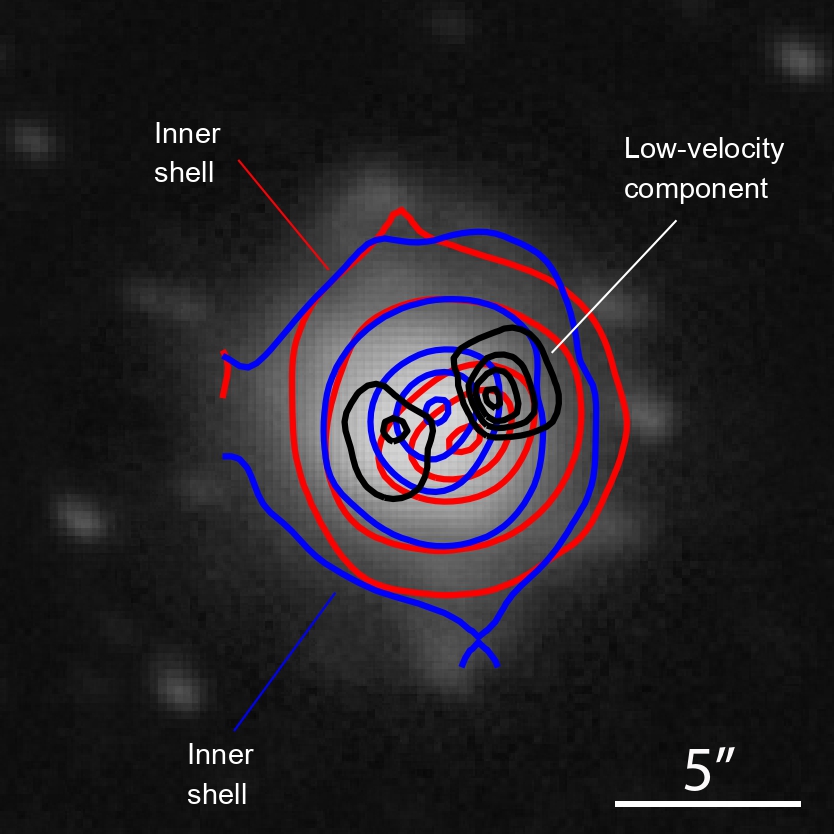}\\
\hspace*{0.015cm}
\includegraphics[width=0.77\linewidth]{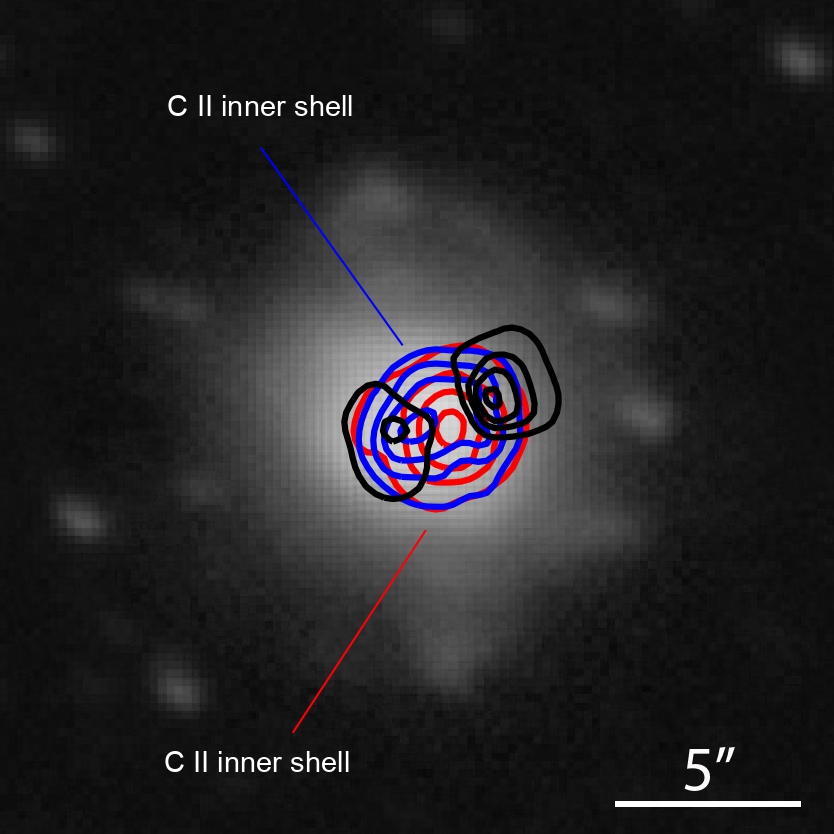}
\caption{Grayscale representation of the colour-composite picture of M\,2-31 shown in Figure~\ref{fig:Componentes}. 
Top and middle figure: GTC MEGARA [N\,{\sc ii}] $\lambda$6584 contours extracted from the flux maps of the five kinematic components presented in Figure~\ref{fig:M2-31_MEGARA-mosaico} are overlaid using different colours. 
Top figure: red and blue for the NE and SW components of the spectroscopic outflow along PA $\simeq$ 50$^\circ$--240$^\circ$, respectively. Middle figure: red and blue for the receding and approaching components of the nebular shells, respectively. Bottom figure: GTC MEGARA  C\,{\sc ii} $\lambda$6578 contours extracted from the flux maps of the two kinematics components presented in Figure~\ref{fig:M2-31_MEGARA-mosaico_CII}, red and blue for the receding and approaching components of the nebular shells. Black for the low-velocity component almost orthogonal to the outflow described above. 
}
\label{fig:SpecOutflows}
\end{center}
\end{figure}

To assess the connection between these kinematic features and the morphological structures of M\,2-31, we present in Figure~\ref{fig:SpecOutflows} grey-scale optical images overlaid with the contours shown in Figures~\ref{fig:M2-31_MEGARA-mosaico} and \ref{fig:M2-31_MEGARA-mosaico_CII}. 
In the top panel of Figure~\ref{fig:SpecOutflows} we show the position of the fast expanding outflows (red and blue contours) and the low-velocity component (black contours) both obtained from [N\,{\sc ii}] $\lambda$6584  normalized flux. 
The Southwest component of the outflow is coincident with the SW knot, implying that it moves at a high velocity with respect to the nebula radial velocity. The outflow extends down to the innermost regions of M\,2-31, peaking at the region in between the maxima of the low-velocity toroidal structure.

In Figure~\ref{fig:SpecOutflows} middle panel we compare the low-velocity component with the contours from the shell of M\,2-31, also obtained from [N\,{\sc ii}] $\lambda$6584 normalized flux. 
Similarly to the fast outflows, the inner shell peaks at locations with no contribution from the low-velocity component.
In Figure~\ref{fig:SpecOutflows} bottom panel we show the position of the contours from the inner shell of M\,2-31 obtained from C\,{\sc ii} $\lambda$6578 normalized flux. 
The contours of the emission in the C\,{\sc ii} $\lambda$6578 line are approximately aligned along the same direction of the contours of the shell of M\,2-31 obtained from the emission in the [N\,{\sc ii}] $\lambda$6584 line, but the C\,{\sc ii} contours are confined within those of [N\,{\sc ii}].

%Contornos

\subsection{The CSPN of M 2-31}

\begin{figure*}
\includegraphics[width=0.95\linewidth, trim=40 25 40 0cm]{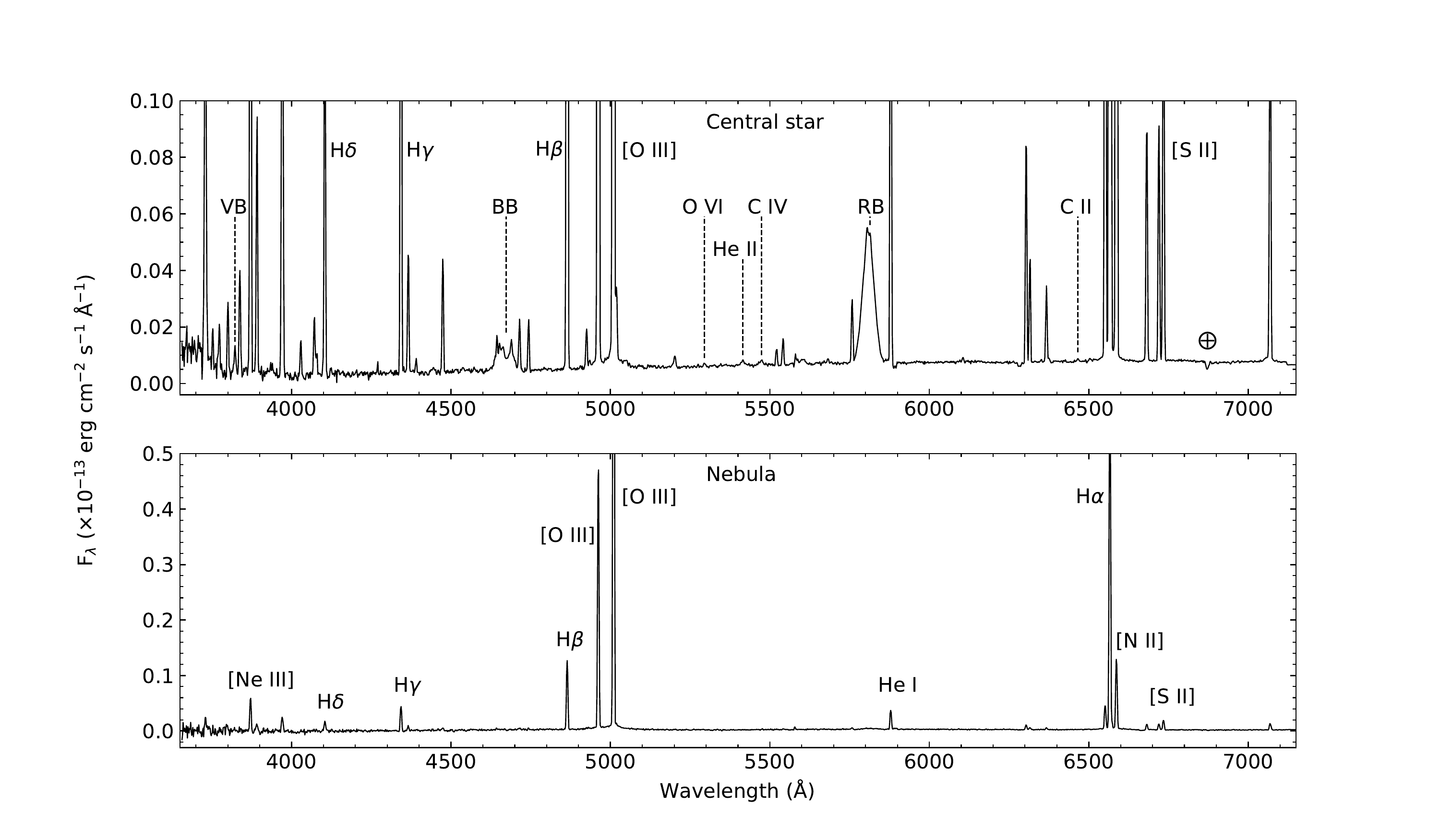}
\caption{One-dimensional NOT ALFOSC spectra of the CSPN (top panel) and nebula (bottom panel) of M\,2-31, as extracted from the apertures A and B shown in the left panel of Figure~\ref{fig:Componentes}, respectively. The VB, BB and RB [WR] features as well as the most prominent emission lines are labeled.}
\label{fig:SpecStarPN}
\end{figure*}

The high-quality and wide spectral coverage of the NOT ALFOSC spectrum of the CSPN of M\,2-31 can be used to assess its spectral classification. The spectrum of its CSPN presented in the top panel of Figure~\ref{fig:SpecStarPN} clearly exhibits the presence of the classic broad WR features known as the blue (BB) and red (RB) bumps located at 4686~\AA\ and 5806~\AA, respectively. 
These broad WR features are known to be composed by the contribution of several emission lines with stellar and nebular origin. For example, the RB can be mostly attributed to the presence of C\,{\sc iv} at 5801~\AA\ and 5812~\AA, while the BB contains contributions from different emission lines, including mainly He\,{\sc ii}, the C\,{\sc iv} broad WR feature and the Ar\,{\sc iv} nebular lines \citep[see][and references therein]{GG2020}.

\label{sec:spec_cspn}

\begin{table*}
\begin{center}
\caption{Parameters for the emission lines obtained from the Gaussian fitting of the WR features of the central star in M\,2-31.}
\label{tab:table_WR}
\begin{tabular}{lcllccccc} 
\hline
ID  &  Nature     & 
\multicolumn{1}{l}{Ion}  & 
\multicolumn{1}{l}{$\lambda_\mathrm{rest}$} & 
\multicolumn{1}{c}{$I_\mathrm{H\beta}$} & 
\multicolumn{1}{c}{FWHM} & 
\multicolumn{1}{c}{EW} & 
\multicolumn{1}{c}{$\lambda_\mathrm{obs}$} & 
\multicolumn{1}{c}{$I_\mathrm{CIV}$} \\
(1) & (2)   & 
\multicolumn{1}{l}{(3)} &  
\multicolumn{1}{l}{(4)} &  
\multicolumn{1}{c}{(5)} &  
\multicolumn{1}{c}{(6)} &  
\multicolumn{1}{c}{(7)} &  
\multicolumn{1}{c}{(8)} &  
\multicolumn{1}{c}{(9)} \\  
\hline
VB &        &               &        &       &       &       &        &     \\
1  &  Neb   &   H10         & 3797.9 & 38.4 $\pm$ 2.6  &  4.8  &  59 $\pm$ 10  & 3800.3 & 19.6 $\pm$ 1.3 \\
2  &  WR    & O~{\sc vi}    & 3822   & 24.1 $\pm$ 2.5  &  8.7  &  44 $\pm$ 8  & 3822.3 & 12.3 $\pm$ 1.3 \\
3  &  Neb   &   H9         & 3835.4 & 57.9 $\pm$ 3.2  &  5.2  & 104 $\pm$ 19 & 3837.7 & 29.6 $\pm$ 1.7 \\
BB &        &               &        &       &       &       &        &     \\
4  &  WR    & C~{\sc ii}    & 4638.9 & 16.3 $\pm$ 0.9  & 20.1  & 23.9 $\pm$ 1.4  & 4641.4 &  8.3 $\pm$ 0.5 \\
5  &  Neb   & O~{\sc ii}+N~{\sc iii}& 4641.6** & 6.0 $\pm$ 0.4   & 3.9   &  8.8 $\pm$ 0.8  & 4644.1 &  3.0 $\pm$ 0.2 \\
6  &  Neb   & O~{\sc ii}*    & 4649.5** & 5.4 $\pm$ 0.5   & 5.4   &  8.0 $\pm$ 0.8  & 4652.0 &  2.7 $\pm$ 0.3 \\
7  &  WR    & C~{\sc iv}    & 4658.5 & 21.1 $\pm$ 0.8  & 14.6  & 31.5 $\pm$ 1.5  & 4662.5 & 10.8 $\pm$ 0.4 \\
8  &  WR    & He~{\sc ii}   & 4685.7 & 28.1 $\pm$ 1.0  & 25.9  & 42.8 $\pm$ 1.8  & 4688.7 & 14.4 $\pm$ 0.5 \\
9  &  Neb   & He~{\sc ii}   & 4685.7 & 4.2 $\pm$ 0.4   & 4.7   &  2.6 $\pm$ 0.3  & 4689.3 &  2.1 $\pm$ 0.2 \\
10  &  Neb   & [Ar~{\sc iv}] & 4711.4 & 17.2 $\pm$ 0.6  & 5.6   & 26.8 $\pm$ 1.5  & 4714.7 &  8.8 $\pm$ 0.3 \\
11  &  Neb   & [Ar~{\sc iv}] & 4740.2 & 15.0 $\pm$ 0.5  & 4.8   & 23.7 $\pm$ 1.4  & 4743.4 &  7.6 $\pm$ 0.3 \\
RB &        &               &        &       &       &       &        &     \\
12 &  Neb   & [N~{\sc ii}]  & 5754.6 & 12.1 $\pm$ 0.3  &  5.0 &  19.7 $\pm$ 0.7  & 5757.8 &  6.2 $\pm$ 0.1 \\
13 &  WR    & C~{\sc iv}    & 5806   & 195.8 $\pm$ 1.0 & 41.7 & 334.2 $\pm$ 3.4  & 5809.2 & 100.0\\ 
14 &  Neb   & He~{\sc i}    & 5875.6 & 129.4 $\pm$ 0.7 &  4.9 & 235 $\pm$ 6  & 5879.1 & 66.1 $\pm$ 0.4 \\
Other&   &   &        &       &       &       &        &     \\
15 &  WR    & O~{\sc vi}    & 5290   &  1.2 $\pm$ 0.2  &  7.8 &   1.8 $\pm$ 0.3  & 5294.9 &  0.6 $\pm$ 0.1 \\
16 &  WR    & He~{\sc ii}   & 5412   &  3.1 $\pm$ 0.3  & 17.2 &   4.6 $\pm$ 0.5 & 5415.2 &  1.6 $\pm$ 0.2 \\
17 &  WR    & C~{\sc iv}    & 5470   &  2.0 $\pm$ 0.2  & 10.8 &   3.0 $\pm$ 0.3  & 5473.7 &  1.1 $\pm$ 0.1 \\
18 &  WR    & C~{\sc ii}    & 6461   &  0.6 $\pm$ 0.2 &  7.3 &   1.1 $\pm$ 0.3  & 6465.9 &  0.3 $\pm$ 0.1 \\
\hline
\end{tabular}
\begin{description}
Columns:
(1) Identification number of the Gaussian components in the WR features.
(2)Nature of the contributing emission line: WR (broad) or nebular (narrow).
(3) Ion responsible for the line.
(4) Rest wavelength in \AA.
(5) Reddening-corrected intensities considering H$\beta$ = 100. %in units of erg cm$^{-2}$ s$^{-1}$ \AA$^{-1}$. 
(6) Full Width at Half Maximum in units of \AA.
(7) Equivalent Width in units of \AA.
(8) Observed center of the line.
(9) Intensities considering $I_\mathrm{CIV}$ = 100.
(*)  O~{\sc ii} multiplet, see in detail in \citet{GarciaRojas2018}.
(**) Average $\lambda_\mathrm{rest}$ of a blend of emission lines that we cannot resolve, but these emission lines can be distinguished in \citet{GarciaRojas2018}. See text for details.
\end{description}
\end{center}
\end{table*}

Other spectral features in the spectrum of the CSPN of M\,2-31 can be attributed to a WR origin whenever their FWHM is larger than the typical FWHM of nebular lines, $\lesssim$5.6~\AA. 
Very notably, the careful inspection of the emission lines in the spectrum of the CSPN has allowed us to identify the O\,{\sc vi} line at 3822~\AA. 
This line, which is referred as the violet bump (VB), is key to classify a CSPN as part of the oxygen sequence. 
Other broad WR features detected in the CSPN spectrum are those of 
O\,{\sc vi}~$\lambda$5290, 
He\,{\sc ii}~$\lambda$5412, 
C\,{\sc iv} $\lambda$5470, and 
C\,{\sc ii}~$\lambda$6461. 
All WR lines are listed in Table~\ref{tab:table_WR}.

\begin{figure*}
\begin{center}
\includegraphics[width=1.0\linewidth, trim= 20 235 20 0cm]{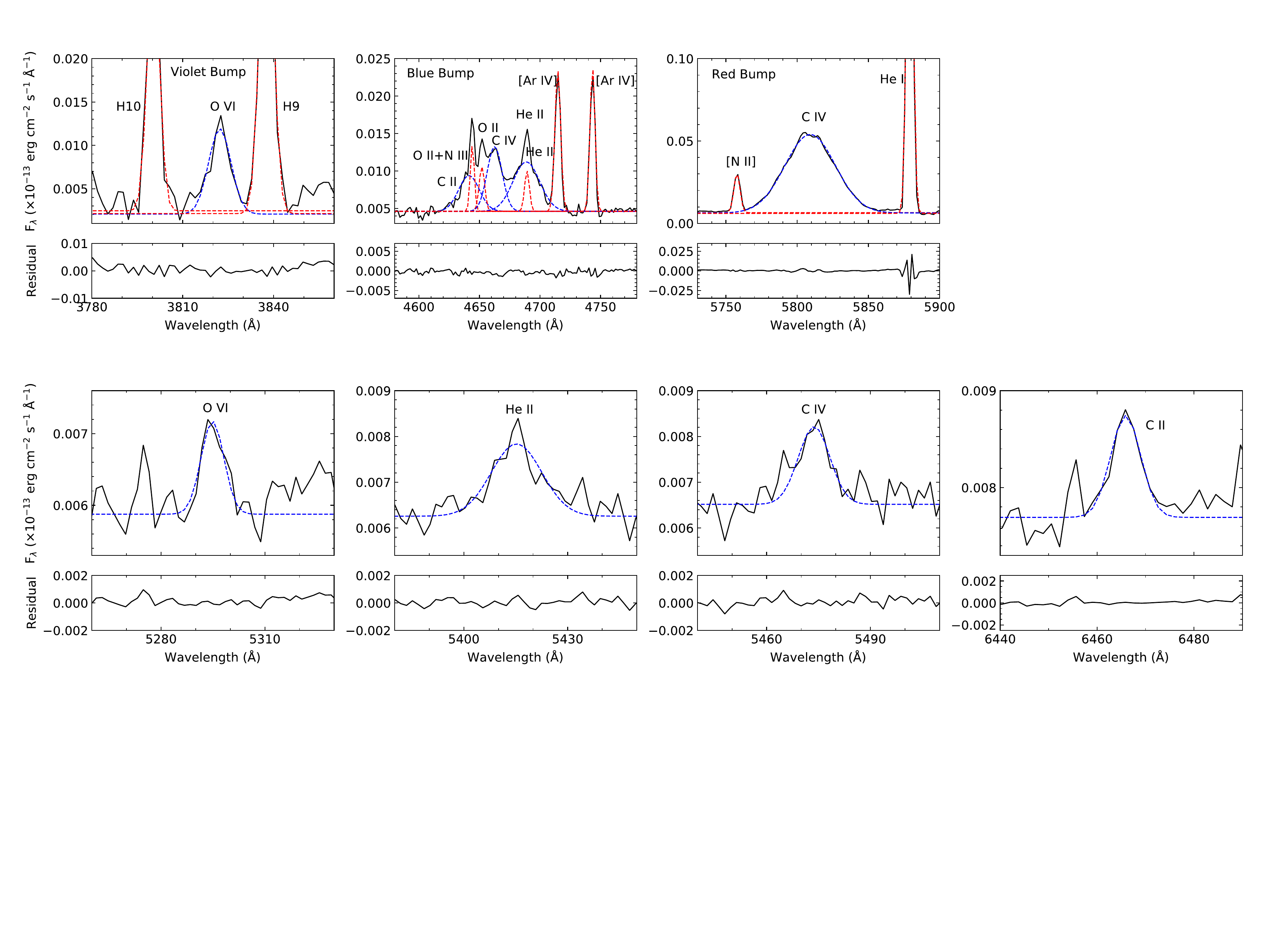}
\caption{Multi-Gaussian fits to the VB (top left), BB (top center) and RB (top right) broad spectral features of the CSPN of M\,2-31. The bottom row show other WR features in the spectrum. 
The blue dashed lines fit the broad (stellar) features, while the red dashed lines fit the nebular emission lines. 
Residuals are shown at the bottom of each panel in units of 10$^{-13}$ erg~cm$^{-2}$~s$^{-1}$~\AA$^{-1}$. 
Details of the fits are listed in Table~\ref{tab:table_WR}.
}
\label{fig:Ajuste_BB_RB}
\end{center}    
\end{figure*}

In order to disentangle the contribution of the multiple nebular and WR lines to the different WR features in the spectrum of the CSPN of M\,2-31, we used the {\sc iraf} task \emph{splot} to deblend those using one Gaussian function for each of them.  
This procedure, illustrated in Figure~\ref{fig:Ajuste_BB_RB}, has allowed us to estimate different line properties, the equivalent widths (EW), FWHMs and observed wavelengths ($\lambda_\mathrm{obs}$) for different contributing lines for the VB, BB, RB and other WR features. 
Each panel in Figure~\ref{fig:Ajuste_BB_RB} shows the residuals of the fit after subtracting the modelled lines. 
The uncertainties in the line intensities and EWs were calculated following the expressions presented by \citet{Tresse1999}:
\begin{equation} 
 \sigma_F = \sigma_c D \sqrt{2N_\mathrm{pix} + \frac{\mathrm{EW}}{D}}, 
\end{equation}
\begin{equation} 
 \sigma_\mathrm{EW} = \frac{\mathrm{EW}}{F} \sigma_c D \sqrt{\frac{\mathrm{EW}}{D} + 2N_\mathrm{pix} + (\mathrm{EW}/D)^2/N_\mathrm{pix}}, 
\end{equation}
\noindent 
respectively, where $\sigma_c$ is the mean standard deviation per pixel of the continuum on each side of the line, $N_\mathrm{pix}$ is the number of pixels under the line, $D$ = 1.7 \AA\,pixel$^{-1}$ is the spectral dispersion, and $F$ is the line's flux. The results of our analysis are summarized in Table~\ref{tab:table_WR}. 

The most complex WR feature is the BB, which is composed by several broad and nebular lines. 
A good fit was achieved adopting three broad WR features from C\,{\sc ii}, C\,{\sc iv} and He\,{\sc ii} and five narrow nebular lines corresponding to He\,{\sc ii}, C\,{\sc ii}, N\,{\sc iii} and [Ar\,{\sc iv}]. 
We have relied on \citet{GarciaRojas2018} and the NIST Atomic Spectra Database Lines\footnote{\url{https://physics.nist.gov/PhysRefData/ASD/lines_form.html}} to identify these emission lines. Two of the nebular lines listed in Table~\ref{tab:table_WR}, those at $\lambda_\mathrm{obs}$ 4644.1 and 4652.0~\AA\ present in the BB, corresponds to blends. 
The former corresponds to a blend of the O\,{\sc ii} $\lambda\lambda$4638.86,4641.8 and N\,{\sc iii} $\lambda\lambda$4640.6,4641.85 emission lines, and the latter to a blend of O\,{\sc ii} nebular lines at 4649.13 and 4650.84~\AA.

On the other hand, the RB is composed by a broad C\,{\sc iv} line with almost negligible contribution from nebular He\,{\sc i} and [N\,{\sc ii}] emission lines. 
The remaining broad WR features do not have any contribution from other emission lines. 
These include the O\,{\sc vi} lines at 3822~\AA\ and 5290~\AA\ lines, first reported in this study, which immediately suggest that the CSPN of M\,2-31 belongs to the [WO] sequence. 
Furthermore, we also identify O\,{\sc vi} $\lambda$5290, He\,{\sc ii} $\lambda$5412, C\,{\sc iv} $\lambda$5470 and C\,{\sc ii} $\lambda$6461. 
These are illustrated in the bottom panels of Figure~\ref{fig:Ajuste_BB_RB} and their details are listed in the bottom part of Table~\ref{tab:table_WR}.

To determine the spectral classification of the CSPN of M\,2-31 we use the scheme defined by \citet{AckerNeiner2003} normalizing all the fluxes of the WR broad lines listed in Table~\ref{tab:table_WR} to that of the C\,{\sc iv} $\lambda$5806 line. 
These relative intensities are listed in the last column of Table~\ref{tab:table_WR} labeled as $I_\mathrm{CIV}$. 
The C\,{\sc iv} $\lambda$5806~\AA\ line has a FWHM of 41.7~\AA, which is consistent with a spectral type between a [WO4] and [WO3] \citep[see table~2 in][]{AckerNeiner2003}. However, the normalized flux of the O\,{\sc vi} $\lambda$3822~\AA\ is 12.3, which rather implies a [WO4] spectral type as also suggested by the normalized fluxes of the He\,{\sc ii} $\lambda$4686 and $\lambda$5412~\AA\ lines. Nevertheless, the normalized fluxes of the O\,{\sc vi} $\lambda$5290 and C\,{\sc iv} $\lambda$5470~\AA\ are at the lower range between [WO4] and [WO4pec].

\subsection{Physical properties and chemical abundances}
\label{sec:spec_neb}

The nebular spectrum of M\,2-31 is presented in the bottom panel of Figure~\ref{fig:SpecStarPN}. 
It exhibits multiple O, Ne, S, N, Cl and Ar forbidden lines as well as H and He recombination lines. 
The complete list of emission lines detected in the nebular spectrum is reported in Table~\ref{tab:table_flux}.

This table lists the correspondent ion with its rest wavelength ($\lambda_\mathrm{rest}$), the observed flux ($F_\mathrm{obs}$) and the extinction corrected intensities ($I_\mathrm{corr}$). $F_\mathrm{obs}$ and $I_\mathrm{corr}$ have been normalized to H$\beta$ = 100. A value of $c$(H$\beta$) = 0.92 $\pm$ 0.03 was derived from the H$\alpha$/H$\beta$ ratio and used to unredden the line fluxes using the extinction law from \citet{Cardelli1989} for $R_\mathrm{V}$ = 3.1.

Different line intensity ratios have been used to determine the physical conditions of the gas in M\,2-31 using the extensively tested code {\sc pyneb} \citep{Luridiana2015}. Electron temperatures determined from the [N\,{\sc ii}] lines and [O\,{\sc iii}] auroral to nebular line intensity ratios were estimated to be $T_\mathrm{e}$([N\,{\sc ii}]) = 12500$\pm$1300~K and $T_\mathrm{e}$([O\,{\sc iii}]) = 10300$\pm$1000~K, respectively. Values of the electron densities $n_\mathrm{e}$ were obtained from the [S\,{\sc ii}] doublet line intensity ratio adopting the two different values of $T_\mathrm{e}$ given above, resulting in 5600$\pm$2000~cm$^{-3}$ and 5200$\pm$1900~cm$^{-3}$ for the $T_\mathrm{e}$([N\,{\sc ii}]) and $T_\mathrm{e}$([O\,{\sc iii}]), respectively. A summary of these values is presented in the top rows of Table~\ref{tab:table_abundances}.

\begin{table}
\begin{center}
\caption{Reddeding-corrected nebular line intensity ratios obtained for the spectrum extracted from aperture B (see the green regions in Fig.~\ref{fig:Componentes} left panel) of M\,2-31.}
\label{tab:table_flux}
\begin{tabular}{clcc} % four columns, alignment for each
\hline
$\lambda_\mathrm{rest}$ & \multicolumn{1}{c}{Ion}  &  $F_\mathrm{obs}$  &  $I_\mathrm{corr}$  \\ 
\hline
3727 & [O~{\sc ii}]    &   17 $\pm$ 9 &  32 $\pm$ 20 \\ 
3869 & [Ne~{\sc iii}]  &   45 $\pm$ 12 &  84 $\pm$ 22 \\ 
3889 & He~{\sc i}      &   11 $\pm$ 7  &  20 $\pm$ 13 \\ 
3968 & [Ne~{\sc iii}]  &   21 $\pm$ 15 &   38 $\pm$ 27 \\ 
4102 & H{$\delta$}     &   15 $\pm$ 6  &   24 $\pm$ 10 \\ 
4341 & H{$\gamma$}     &   35 $\pm$ 9  &   49 $\pm$ 13 \\ 
4363 & [O~{\sc iii}]   & 6.0 $\pm$ 1.8 &    8.2 $\pm$  2.4 \\ 
4471 & He~{\sc i}      & 5.7 $\pm$ 2.4 &    7.3 $\pm$  3.0 \\ 
4740 & [Ar~{\sc iv}]   & 2.1 $\pm$ 0.7 &    2.2 $\pm$  0.8 \\ 
4861 & H{$\beta$}      &     100       &  100 \\ 
4959 & [O~{\sc iii}]   &  392 $\pm$12 &  371 $\pm$ 11 \\ 
5007 & [O~{\sc iii}]   & 1195 $\pm$26 & 1103 $\pm$ 21 \\ 
5200 & [N~{\sc i}]     & 2.2 $\pm$ 0.1 &    1.9 $\pm$  0.7 \\ 
5538 & [Cl~{\sc iii}]  & 1.1 $\pm$ 0.5 &    0.79 $\pm$  0.33 \\ 
5755 & [N~{\sc ii}]    & 2.4 $\pm$ 0.5 &    1.65 $\pm$  0.34 \\ 
5876 & He~{\sc i}      & 27.9 $\pm$ 1.1 &   18.1 $\pm$  0.6 \\ 
6300 & [O~{\sc i}]     & 7.9 $\pm$ 0.6 &    4.52 $\pm$  0.30 \\ 
6312 & [S~{\sc iii}]   & 4.1 $\pm$ 0.5 &    2.35 $\pm$  0.27 \\ 
6364 & [O~{\sc i}]     & 2.8 $\pm$ 0.4 &    1.59 $\pm$  0.21 \\ 
6548 & [N~{\sc ii}]    & 38.1 $\pm$ 1.6 &   20.3 $\pm$  0.8 \\ 
6563 & H{$\alpha$}     &  536 $\pm$12 &  285 $\pm$ 6\\ 
6584 & [N~{\sc ii}]    & 109.1 $\pm$ 3.2 &   57.7 $\pm$  1.3 \\ 
6678 & He~{\sc i}      & 7.9 $\pm$ 0.9 &    4.1 $\pm$  0.5 \\ 
6716 & [S~{\sc ii}]    & 8.4 $\pm$ 0.5 &    4.26 $\pm$  0.25 \\ 
6731 & [S~{\sc ii}]    & 14.1 $\pm$ 0.7 &    7.14 $\pm$  0.30 \\ 
7065 & He~{\sc i}      & 10.3 $\pm$ 0.5 &    4.77 $\pm$  0.22 \\ 
\hline
$c$(H$\beta$)&            &0.92$\pm$0.03      &                        \\ 
\hline
log($F$(H$\beta$))&      &   $-$13.14$\pm$0.01&                        \\ 
%AV   &  1.98$\pm$0.07  &                    &                   \\ 
\hline
\end{tabular}
\end{center}
\end{table}

We also used {\sc pyneb} to calculate ionic abundances taking into account the aforementioned physical conditions. 
These are listed in Table~\ref{tab:table_abundances} middle rows. 
The total chemical abundances of M\,2-31, listed in the bottom part of Table~\ref{tab:table_abundances}, were derived adopting the ionization correction factor (ICFs) provided by \citet{Delgado-Inglada2014}, but for the N abundances, computed using the ICF derived by \citet{Kingsburgh1994}. We discuss these results in the next section. The error propagation on the emission lines and other parameters (such as temperature, density, and abundances)  was obtained using a Monte Carlo procedure implemented within {\sc pyneb}. Details will be presented in a forthcoming paper (Sabin et al. in prep.).

\begin{table}
\begin{center}
\caption{Physical properties, ionic and total abundances for the region B in M\,2-31.}
\label{tab:table_abundances}
\begin{tabular}{lcc}
\hline
Parameter &  Condition &  Value  \\
\hline
%\midrule
$T_\mathrm{e}$([N\,{\sc ii}]) [K]  &  $n_\mathrm{e}$([S~{\sc ii}])&  12500 $\pm$ 1300 \\
$T_\mathrm{e}$([O\,{\sc iii}]) [K] &  $n_\mathrm{e}$([S~{\sc ii}])&  10300 $\pm$ 1000 \\
$n_\mathrm{e}$([S~{\sc ii}]) [cm$^{-3}$] &  $T_\mathrm{e}$([N~{\sc ii}]) & 5600 $\pm$ 2000 \\
$n_\mathrm{e}$([S~{\sc ii}]) [cm$^{-3}$] &  $T_\mathrm{e}$([O~{\sc ii}]) & 5200 $\pm$ 1900 \\
%\toprule
\hline
X/H$^+$                 &                    & ionic abundance\\
\hline
He$^{+}$/H$^{+}$  & & 0.111 $\pm$ 0.022 \\
N$^{0}$/H$^{+}$   & & (5.3 $\pm$ 3.3) $\times 10^{-6}$ \\
N$^{+}$/H$^{+}$   & & (7.7 $\pm$ 2.1) $\times 10^{-6}$ \\
O$^{0}$/H$^{+}$   & & (4.6 $\pm$ 1.9) $\times 10^{-6}$ \\
O$^{+}$/H$^{+}$   & & (1.1 $\pm$ 0.8) $\times 10^{-5}$ \\
O$^{+2}$/H$^{+}$  & & (3.6 $\pm$ 0.2) $\times 10^{-4}$ \\
Ne$^{+2}$/H$^{+}$ & & (9.8 $\pm$ 3.0) $\times 10^{-5}$ \\
S$^{+}$/H$^{+}$   & & (3.2 $\pm$ 1.0) $\times 10^{-7}$ \\
S$^{+2}$/H$^{+}$  & & (2.2 $\pm$ 1.2) $\times 10^{-6}$ \\
Cl$^{+2}$/H$^{+}$ & & (4.9 $\pm$ 2.9) $\times 10^{-8}$ \\
Ar$^{+3}$/H$^{+}$ & & ~~~~~(6 $\pm$ 4) $\times 10^{-7}$ \\
\hline
Element & & total abundance  \\
\hline
He/H & & 0.111 $\pm$ 0.022 \\
N/H  & & (3.3 $\pm$ 1.9) $\times 10^{-4}$ \\
O/H  & & (3.7 $\pm$ 1.2) $\times 10^{-4}$ \\
Ne/H & & (1.2 $\pm$ 0.6) $\times 10^{-4}$ \\
S/H  & & (5.9 $\pm$ 2.2) $\times 10^{-6}$ \\
Cl/H & & (1.1 $\pm$ 0.5) $\times 10^{-7}$ \\
\hline
\end{tabular}
\end{center}
\end{table}

\section{Discussion}

In this paper we present a study of M\,2-31 using information obtained from optical images and medium-resolution long-slit and high-dispersion IFS observations. 
Combining all this information we have been able to unravel the main morpho-kinematic components of this PN. 
M\,2-31 is found to have a complex morphology, showing multiple components with some of them more evident in direct images, but other revealed only by the unprecedented tomographic capabilities of the MEGARA IFU observations. 
In the following we discuss different aspects of M\,2-31 regarding the nebula and its CSPN.

\subsection{Morphology and kinematics of M\,2-31}

The line-fitting procedure of the MEGARA observations presented in Section~\ref{sec:multi-component} disclosed the presence of a number of morphological components with specific kinematic signatures. 
These can be interpreted as three main components:
\begin{enumerate}
\item 
A pair of fast spectroscopic outflows expanding along PA $\approx$ 50$^\circ$ and PA $\approx$ 240$^\circ$ (first and last columns of Fig.~\ref{fig:M2-31_MEGARA-mosaico}). 
These outflows seem to present a homologous expansion, i.e.\ their velocity increases with the distance to the CSPN, as shown in Figure~\ref{fig:M2-31_MEGARA-mosaico} where their tips have  velocity values higher by $\simeq$33~km~s$^{-1}$ than those of the regions close to the center of M\,2-31. The averaged expansion velocity of these outflows with respect to the CSPN is 63.8~km~s$^{-1}$. The SW knot is found to be associated with the tip of the blue fast outflow expanding towards that direction. These spectroscopic outflows, which are only revealed thanks to MEGARA's tomographic capabilities, make M\,2-31 a member of the spectroscopic bipolar nebulae group.

\item 
An inner cavity that accounts for the brightest optical emission from the inner regions and most of the extended IR emission (second and fourth columns of Fig.~\ref{fig:M2-31_MEGARA-mosaico}). 
This shell has an expansion velocity of 21.3~km~s$^{-1}$ with respect to the CSPN, with a subtle velocity gradient more or less oriented with that of the fast outflow.

\item
A faint outer shell with similar expansion velocity than the inner shell.  

\item
A pair of knots along PA = 10$^{\circ}$ that are associated with the N and S knots disclosed in the optical and IR images of M\,2-31 (Fig.~\ref{fig:Componentes}). 
These features do not exhibit strong kinematic signatures and seem to be embedded within the outer shell. 
One might think of them as the low-ionization dense clumps commonly find in the outer shells of multiple shell PNe \citep[see, e.g.,][]{Goncalves2009,GD2012}.

\item 
A low-velocity component orthogonally-oriented to the fast spectroscopic outflows (central column of Fig.~\ref{fig:M2-31_MEGARA-mosaico}). 
This is a slowly expanding structure with a velocity $\sim$8~km~s$^{-1}$ with respect to the CSPN. 
Its velocity structure is suggestive of a toroidal structure surrounding the CSPN. 
The lack of emission coming from the central regions might be attributed to the line-fitting procedure, as the emission from this component is overcome by the much brighter emission from the extended shell. 
Alternatively, this slowly expanding structure could be attributed to bipolar lobes or an outflow along this direction that would have a small inclination angle with respect to the plane of the sky. 
% might be tracing a waist-like structure surrounding the CSPN.

\end{enumerate}

Our interpretation of the MEGARA data of M\,2-31 suggests that its symmetry axis is aligned with the fast outflows at PA $\approx 50^{\circ}$ rather than along PA $\approx 10^{\circ}$ with the N and S knots.  
Some of the structures unveiled by our MEGARA data were first described by \citet{AkrasLopez2012}. 
These authors presented the analysis of high-dispersion long-slit SPM MES observations positioned at PA = 0$^{\circ}$ and 35$^{\circ}$ and showed the presence of the inner cavity in M\,2-31 associated to the waist-like structure surrounding the CSPN. 
Although none of their slits were oriented with the symmetry axis of M\,2-31 (PA $\approx$ 50$^{\circ}$), they suggested it had to be close to 35$^{\circ}$.

The kinematical information of M\,2-31 gathered so far allows us to envisage its true physical structure. 
We propose that M\,2-31 has a very similar physical structure as that proposed by \citet{AkrasLopez2012} for a different PN, namely M\,1-32 (see the leftmost column of their Figure~3). 
From inside-out: 
i) an inner elongated cavity surrounded by a low-velocity pinched waist, 
ii) a bipolar structure aligned with the symmetry axis, which for M\,2-31 corresponds to the fast outflows along PA $50^\circ$, 
and 
iii) an extended slow outer shell.

\subsection{Reclassification of the spectral type of the CSPN of M\,2-31}

A wealth of WR spectral features is detected in the NOT ALFOSC spectrum of the CSPN of M\,2-31, including some previously undetected such as the O\,{\sc vi} at 3822~\AA\ (the VB) and O\,{\sc vi} 5290~\AA\, broad lines.  
The detection of these features immediately hints at this CSPN belonging to the [WO] sequence.  
A multi-Gaussian fitting procedure has provided as with comprehensive quantitative information of the different WR features in the spectrum of the CSPN of M\,2-31, including the complex BB around $\sim$4650~\AA\ that includes contributions of O\,{\sc ii} and N\,{\sc iii} nebular lines at 4640~\AA\ and 4665~\AA\ as shown in the higher resolution spectra presented by \citet{GarciaRojas2018} in the bottom right panel of their Figure~4.

The identification of the O~{\sc vi} lines and the normalized intensities of the [WR] features with respect to C~{\sc iv} $\lambda$5806 line indicates that the CSPN of M\,2-31 spectrum seems consistent with a [WO4] spectral type in the \citet{AckerNeiner2003} scheme. 
This classification differs from previous [WC4-6] classifications \citep{Tylenda1993,AckerNeiner2003}, most likely due to the low quality of their spectroscopic observations caused by the weakness of the CSPN of M\,2-31.

\subsection{M\,2-31 from chemical point of view}

The physical conditions and the chemical abundances of M\,2-31 were calculated from the NOT ALFOSC long-slit spectroscopic observations obtained along a PA = 10$^\circ$ (see Tab.~\ref{tab:table_abundances}). 
The He/H ratio is $\simeq$0.11, consistent with those of Type II PNe, but the N/O abundances ratio $\simeq$0.7 is rather suggestive of a Type~I origin \citep[][]{Peimbert1978}\footnote{The limits between Type I and II are He/H=0.125 and N/O=0.5.}. 
We note that the N/O increases up to $\simeq$4 if the ICF for N provided by \citet{Delgado-Inglada2014} were to be used, but those authors already noticed that such ICF does not apply properly to the case of M\,2-31.

Comparing our abundance determinations with those of the most recent study of \citet{GarciaRojas2018}, the He and Cl abundances agree relatively good, within 0.2 dex, whereas those of N, O and S differ by more than 0.4 dex. 
The N abundances and N/O abundances ratio show the most notable differences, with much higher values in our study.  
Indeed, the N/O ratio reported by \citet{GarciaRojas2018}, N/O = 0.38, suggests a Type~II PN, whereas the value of this ratio reported by \citet{DelgadoInglada2015}, N/O = 0.68, is consistent with ours. 
The lower N/O ratio proposed by \citet{GarciaRojas2018} could be partially attributed to the different location of the apertures onto the nebula used for spectral extraction, but it is certainly governed by the lower value of the extinction correction used by those authors compared to ours, which reduces their electronic temperature and more importantly their extinction correction to the blue [O~{\sc ii}] $\lambda$3727 emission line used to determine the O$^+$/H$^+$ abundances and O/N abundances ratio.

\subsection{On the spatio-kinematics of the C\,{\sc ii} emission line}

The MEGARA observations of M\,2-31 have been used to produce a 2D map of the spatial distribution of the C\,{\sc ii} $\lambda$6578 emission line for the first time in a PN. 
The emission of this C~{\sc ii} line is associated with two kinematical components that are confined within the innermost regions of M\,2-31 with a radius of $\sim$2~arcsec. 
The expansion velocity implied for these two components is 16.3~km~s$^{-1}$.

This emission can be compared with the brightest emission in the [N\,{\sc ii}] $\lambda$6584 emission line, with a radius of 5~arcsec and an expansion velocity of 21.3~km~s$^{-1}$.  
The structure traced by the C\,{\sc ii} emission line is thus enclosed by that traced by the [N\,{\sc ii}] emission and expands slower. 
These results are consistent to those presented by \citet{Richer2017} using long-slit, high-resolution spectroscopic observations obtained with the SPM MES for a sample of 76 PNe. 
Such behaviour is in sharp contrast with the expectations for a chemically homogeneous nebular plasma in ionization equilibrium. Therefore, these differences are more likely produced by spatially-segregated multiple plasma components \citep[e.g., as in NGC\,6778;][]{GarciaRojas2016}, which are important when calculating chemical abundances from permitted and collisionally excited emission lines. Although it was suggested in the past that the C\,{\sc ii} emission line might be excited indirectly by fluorescence \citep[see][]{Grandi1976}, \citet{Richer2017} showed that such physical process never dominates the total surface brightness of this emission line.

M\,2-31 has been classified as a mixed-chemistry dust (MCD) PN, which are objects with spectral features revealing both C-rich (such as PAHs) and O-rich dust (e.g., silicates) \citep[see][]{DelgadoIngladaRodriguez2014,DelgadoInglada2015}. 
Such objects have been proven difficult to explain, with a few mechanisms able to produce (or deplete) C or O in PNe during their formation and evolution which are also metallicity-dependent \citep[see, e.g.,][]{PereaCalderon2009}. To peer into this problem, \citet{GarciaRojas2018} investigated the chemical abundances and specifically the C/O ratios of a sample of MCD PNe using high-resolution spectrocopic observations targeting optical recombination lines from C\,{\sc ii} and O\,{\sc ii}. According to them, most MCD PNe, including M\,2-31, are O-rich and have C/O$<$1, suggesting that they might be the result of the evolution of intermediate-mass stars in the low ($M_\mathrm{ZAMS}<$1.5~M$_{\odot}$) or high ($M_\mathrm{ZAMS}>$3~M$_{\odot}$) mass regimes. If this were the case, PAHs would form as the result of the evaporation of a CO torus \citep[e.g.,][]{Guzman-Ramirez2014}.

Our GTC MEGARA observations presented here are indeed suggestive of the presence of a toroidal inner structure in M\,2-31.
Future IFS observations of M\,2-31 targeting the O~{\sc ii} 4650~\AA\, optical recombination line with the medium-resolution Volume Phased Holographic VPH481-MR ($R\sim$12,000) can help derive more accurate C/O abundance ratios for different kinematic structures in this PN. 
Spatially resolved studies at high spectral resolution confirm GTC MEGARA as a key instrument to peer into the abundance discrepancy problem in ionized nebulae and the MCD phenomenon in PNe \citep{GarciaRojas2016}.

%%%%%%%%%%%%%%%%%%%%%
%%%%%%%%%%%%%%%%%%%%%

\section{Summary}

We presented a comprehensive study of the morphology and kinematics of M\,2-31 using NOT ALFOSC narrow-band images and intermediate-dispersion long-slit spectra and GTC MEGARA high-dispersion IFS observations. 
The following is a summary of our findings:

\begin{itemize}

\item 
The kinematics of PNe is classically investigated using PV maps obtained from long-slit high-dispersion spectra. 
Keeping this technique, a deep analysis of the morpho-kinematic structure of M\,2-31 was carried out extracting PV maps along pseudo-slits at different PAs from the MEGARA data cube. In addition, a sophisticated analysis was made using 2D flux, expansion velocity and FWHM maps of the [N\,{\sc ii}] emission line in combination with NOT ALFOSC image.  It can be concluded that M\,2-31 has five principal kinematic components: 
(i) a pair of fast (expansion velocity $\pm$63.8~km~s$^{-1}$) spectroscopic outflows, which makes M\,2-31 a member of the spectroscopic bipolar nebulae; 
(ii and iii) receding and approaching components of a bright inner and a faint outer shells with an expansion velocity of 21.3~km~s$^{-1}$;  
(iv) a low-velocity component, almost orthogonal to the spectroscopic outflows, with an expansion velocity $\pm$8.1~km~s$^{-1}$; and 
(v) two knots at North and South.

\item
This is the first time that flux, velocity and FWHM 2D maps of the C~{\sc ii} $\lambda$6578 emission line are presented for a PN. 
We found that this permitted line is more concentrated in the central region, with smaller diameter and expansion velocity than the emission of [N\,{\sc ii}].

\item
The spectral properties of the CSPN of M\,2-31 were studied using multi-Gaussian fits to determine the real contribution of the WR spectral lines and the spectral sub-type of the CSPN. 
Considering the WR lines detected in the NOT ALFOSC spectrum and their FWHM and $I_\mathrm{CIV}$, the CSPN of M\,2-31 is reassigned a spectral type [WO4].

\item
The physical properties and the chemical abundances of M\,2-31 have been computed. 
According to the Helium abundances, M\,2-31 is a type~II PN, but according to the N abundances and N/O ratio it is rather a type~I PN. 

\end{itemize}

\section*{Acknowledgements}

The authors would like to thank the referee for a critical reading of the manuscript which helped clarify important aspects.
JSRG and VMAGG acknowledge support from the Programas de Beca Posdoctorales funded by Direcci\'on General de Asuntos del Personal Acad\'emico (DGAPA) of the Universidad Nacional Aut\'onoma de M\'exico (UNAM). 
JAT acknowledges funding by DGAPA UNAM PAPIIT project IA100720 and the Marcos Moshinsky Foundation (Mexico). 
SC and MAG acknowledges financial support from State Agency for Research of the Spanish MCIU through the "Center of Excellence Severo Ochoa" award to the Instituto de Astrof\'isica de Andaluc\'ia (SEV-2017-0709). 
MAG acknowledges support of the Spanish Ministerio de Ciencia, Innovaci\'on y Universidades (MCIU) grant PGC2018-102184-B-100. LS acknowledges support from PAPIIT-UNAM grant IN101819. GR-L acknowledges support from CONACyT (grant 263373) and PRODEP (Mexico). 
The GTC Science Operations group is recognized for scheduling the GTC MEGARA observations beneath the exacting conditions requested by this program.

\section*{DATA AVAILABILITY}

The data underlying this article will be shared on reasonable request to the corresponding author.



\bibliographystyle{mnras}
\bibliography{M2_31_MNRAS} % if your bibtex file is called example.bib

%%%%%%%%%%%%%%%%%%%%%%%%%%%%%%%%%%%%%%%%%%%%%%%%%%

% Don't change these lines
\bsp	% typesetting comment
\label{lastpage}
\end{document}